\documentclass[twocolumn]{aastex631}

\usepackage{comment}
\usepackage{amsmath}
\usepackage{amssymb}
\graphicspath{{./}{Figures/}}

\begin{document}

\title{SILVERRUSH. XIII. 
A Catalog of 20,567 Ly$\alpha$ Emitters at $z=2-7$ Identified\\
in the Full-depth Data of the Subaru/HSC-SSP and CHORUS Surveys}

\correspondingauthor{Satoshi Kikuta}
\email{kikuta.astro@gmail.com}

\author[0000-0003-3214-9128]{Satoshi Kikuta}
\affiliation{National Astronomical Observatory of Japan,
2-21-1, Osawa, Mitaka, Tokyo 181-8588, Japan}
\author[0000-0002-1049-6658]{Masami Ouchi}
\affiliation{National Astronomical Observatory of Japan,
2-21-1, Osawa, Mitaka, Tokyo 181-8588, Japan}
\affiliation{Institute for Cosmic Ray Research, The University of Tokyo,
5-1-5 Kashiwanoha, Kashiwa, Chiba 277-8582, Japan}
\affiliation{Graduate University for Advanced Studies (SOKENDAI), 2-21-1 Osawa, Mitaka, Tokyo 181-8588, Japan}
\affiliation{Kavli Institute for the Physics and Mathematics of the Universe (WPI), University of Tokyo, Kashiwa, Chiba 277-8583, Japan}
\author{Takatoshi Shibuya}
\affiliation{Kitami Institute of Technology, 165 Koen-cho, Kitami, Hokkaido 090-8507, Japan}
\author[0000-0002-2725-302X]{Yongming Liang}
\affiliation{Institute for Cosmic Ray Research, The University of Tokyo,
5-1-5 Kashiwanoha, Kashiwa, Chiba 277-8582, Japan}
\affiliation{National Astronomical Observatory of Japan,
2-21-1, Osawa, Mitaka, Tokyo 181-8588, Japan}
\author{Hiroya Umeda}
\affiliation{Institute for Cosmic Ray Research, The University of Tokyo,
5-1-5 Kashiwanoha, Kashiwa, Chiba 277-8582, Japan}
\affiliation{Department of Physics, Graduate School of Science, The University of Tokyo,
7-3-1 Hongo, Bunkyo, Tokyo 113-0033, Japan}
\author{Akinori Matsumoto}
\affiliation{Institute for Cosmic Ray Research, The University of Tokyo,
5-1-5 Kashiwanoha, Kashiwa, Chiba 277-8582, Japan}
\affiliation{Department of Physics, Graduate School of Science, The University of Tokyo,
7-3-1 Hongo, Bunkyo, Tokyo 113-0033, Japan}
\author[0000-0002-2597-2231]{Kazuhiro Shimasaku}
\affiliation{Department of Astronomy, School of Science, The University of Tokyo,
7-3-1 Hongo, Bunkyo-ku, Tokyo 113-0033, Japan}
\affiliation{Research Center for the Early Universe, School of Science,
The University of Tokyo, 7-3-1 Hongo, Bunkyo-ku, Tokyo 113-0033, Japan}
\author[0000-0002-6047-430X]{Yuichi Harikane}
\affiliation{Institute for Cosmic Ray Research, The University of Tokyo,
5-1-5 Kashiwanoha, Kashiwa, Chiba 277-8582, Japan}
\author[0000-0001-9011-7605]{Yoshiaki Ono}
\affiliation{Institute for Cosmic Ray Research, The University of Tokyo,
5-1-5 Kashiwanoha, Kashiwa, Chiba 277-8582, Japan}
\author[0000-0002-7779-8677]{Akio K. Inoue}
\affiliation{Waseda Research Institute for Science and Engineering, Faculty of Science and Engineering, Waseda University, 3-4-1 Okubo, Shinjuku, Tokyo 169-8555, Japan}
\affiliation{Department of Physics, School of Advanced Science and Engineering, Faculty of Science and Engineering, Waseda University, 3-4-1 Okubo, Shinjuku, Tokyo 169-8555, Japan}
\author{Satoshi Yamanaka}
\affiliation{General Education Department, National Institute of Technology, Toba College,
1-1, Ikegami-cho, Toba, Mie 517-8501, Japan}
\author[0000-0002-3801-434X]{Haruka Kusakabe}
\affiliation{National Astronomical Observatory of Japan,
2-21-1, Osawa, Mitaka, Tokyo 181-8588, Japan}
\affiliation{Observatoire de Gen\`eve, Universit\'e de Gen\`eve, 51 Chemin de P\'egase, 1290 Versoix, Switzerland}
\author[0000-0002-8857-2905]{Rieko Momose}
\affiliation{Carnegie Observatories, 813 Santa Barbara Street, Pasadena, CA 91101, USA}
\affiliation{Department of Astronomy, School of Science, The University of Tokyo,
7-3-1 Hongo, Bunkyo-ku, Tokyo 113-0033, Japan}
\author[0000-0003-3954-4219]{Nobunari Kashikawa}
\affiliation{Department of Astronomy, School of Science, The University of Tokyo,
7-3-1 Hongo, Bunkyo-ku, Tokyo 113-0033, Japan}
\affiliation{Research Center for the Early Universe, School of Science,
The University of Tokyo, 7-3-1 Hongo, Bunkyo-ku, Tokyo 113-0033, Japan}
\author[0000-0003-1747-2891]{Yuichi Matsuda}
\affiliation{National Astronomical Observatory of Japan,
2-21-1, Osawa, Mitaka, Tokyo 181-8588, Japan}
\affiliation{Graduate University for Advanced Studies (SOKENDAI), 2-21-1 Osawa, Mitaka, Tokyo 181-8588, Japan}
\affiliation{Cahill Center for Astronomy and Astrophysics, California Institute of Technology,
MS 249-17, Pasadena, CA 91125, USA}
\author[0000-0003-1700-5740]{Chien-Hsiu Lee}
\affiliation{NSF’s National Optical-Infrared Astronomy Research Laboratory, 950 N Cherry Ave., Tucson, AZ 86719, USA}



\begin{abstract}
We present 20,567 Ly$\alpha$ emitters (LAEs) at $z=2.2-7.3$ that are photometrically identified by the SILVERRUSH program in a large survey area up to 25 deg$^2$ with deep images of five broadband filters (grizy) and seven narrowband filters targeting Ly$\alpha$ lines at $z=2.2$, $3.3$, $4.9$, $5.7$, $6.6$, $7.0$, and $7.3$ taken by the Hyper Suprime-Cam Subaru Strategic Program (HSC-SSP) and the Cosmic HydrOgen Reionization Unveiled with Subaru (CHORUS) survey. 
We select secure $>5\sigma$ sources showing narrowband color excesses via Ly$\alpha$ break screening, taking into account the spatial inhomogeneity of limiting magnitudes. 
After removing spurious sources by careful masking and visual inspection of coadded and multiepoch images obtained over the 7 yr of the surveys, we construct LAE samples consisting of 
6995, 4641, 726, 6124, 2058, 18, and 5 LAEs at $z=2.2$, 3.3, 4.9, 5.7, 6.6, 7.0, and 7.3, respectively, although the $z=7.3$ candidates are tentative. 
Our LAE catalogs contain 289 spectroscopically confirmed LAEs at the expected redshifts from previous work. 
We demonstrate that the number counts of our LAEs are consistent with previous studies with similar LAE selection criteria. 
The LAE catalogs will be made public on our project webpage with detailed descriptions of the content and ancillary information about the masks and limiting magnitudes.
\end{abstract}



\section{Introduction} \label{sec:intro}
Recent deep optical imaging observations for high-redshift galaxies 
step into the new regime in survey parameter space.
One of the most significant advancements is the survey area
that exceeds $\sim 10$ deg$^2$, as achieved by the
8m Subaru Hyper Suprime-Cam (HSC) survey \citep{Aihara2018} 
and the Dark Energy Survey \citep[DES;][]{Abbott2018}.
These large-area data provide statistical measurements
for unexplored low-number-density objects with negligible statistical errors, 
such as the luminosity function at the galaxy-active galactic nucleus (AGN) transition magnitudes
\citep{Ono2018,Ono2021,Ouchi2018,Harikane2018a,Harikane2022,Konno2018,Stevans2018,Guarnieri2019,Adams2020,Moutard2020, Taylor2021}. 
Moreover, large-area surveys have identified
rare objects including high-$z$ Ly$\alpha$ blobs \citep[e.g.,][]{Shibuya2018laelab,Kikuta2019,Taylor2020,Zhang2021c} and protoclusters \citep[e.g.,][]{Toshikawa2018,Harikane2019,Hu2021}.

Large-area imaging surveys are specifically useful for 
studying Ly$\alpha$ emitters (LAEs), which are probes of cosmic reionization
and galaxy formation \citep{Ouchi2020}.
The large volume is necessary to constrain
the cosmological average properties of galaxy Ly$\alpha$ emission 
escaping from the interstellar and intergalactic media.
Large-area observations for LAEs are complementary to
small volume but very deep spectroscopic studies with
the Very Large Telescope/MUSE \citep{Bacon2023} that reveal spectroscopic
properties including Ly$\alpha$ spectral and spatial profiles \citep{Leclercq2017,Leclercq2020a}
down to very faint luminosity limits 
of $\sim 10^{41}-10^{42}$ erg s$^{-1}$ at $z=3-7$.

One of the major large-area LAE surveys is the Hyper Suprime-Cam Subaru Strategic Program \citep[HSC-SSP;][]{Aihara2018,Aihara2018a,Aihara2019,Aihara2022}. 
The survey takes narrowband (NB) images covering up to a total area of $25$ deg$^2$ using four NB filters, NB387, NB816, NB921, and NB1010,
whose central wavelengths are 
3862, 8177, 9215, and 10097 \AA\ targeting
LAEs at $z=2.2$, $5.7$, $6.6$, and $7.3$,
respectively, by NB excess \citep{Ouchi2018} 
with broadband (BB) images taken with five BB filters 
(g, r, i, z, and y bands). The HSC-SSP observations started in 2014 and were completed 
at the end of 2021. 
The completed NB and BB imaging data are reduced and released in the S21A internal release (and will be made public in the final data release, Public Data Release 4). 
In the COSMOS field that is part of the HSC-SSP survey imaging footprints, deep imaging observations with additional HSC NB filters were conducted in the Subaru intensive program entitled 
the Cosmic HydrOgen Reionization Unveiled with 
Subaru \citep[CHORUS;][]{Inoue2020}.
The CHORUS program has taken deep imaging data
with the NB387 filter as well as 
the NB527, NB718, and NB973 filters whose
central wavelengths are 5260, 7170, and 9713 \AA\ 
tracing Ly$\alpha$ emission of LAEs 
at $z=3.3$, $4.9$, and $7.0$, respectively.
In the past, LAE catalogs were made with the HSC-SSP NB images \citep{Shibuya2018laelab}, the CHORUS NB images \citep{Itoh2018,Zhang2020b},
and both of the NB images \citep{Ono2021}.
However, the HSC-SSP imaging observations were not fully completed at the time of these cataloging efforts. 
Part of the HSC-SSP NB imaging was missing even in \citet{Ono2021}, and all of these publications
did not use the HSC-SSP full-depth BB images for the offbands of
the LAE selections.

\begin{figure}[]
    \includegraphics[width=\linewidth]{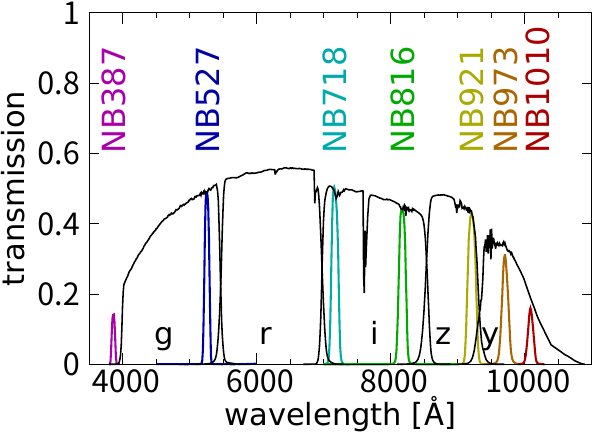}
    \caption{Total transmission curves of relevant HSC filters based on the area-weighted mean measurements, including the transmission of the atmosphere, the prime focus unit, and the dewer window; the primary mirror reflectivity; average CCD quantum efficiency; and the effect of vignetting. Colored thick curves show the transmission of NB filters, from left to right: NB387, NB527, NB718, NB816, NB921, NB973, and NB1010. Black curves show the transmission of HSC broadband filters, from left to right: g, r2, i2, z, and y. Note that the r- and i-band filters were replaced with the new r2 and i2 filters in 2016, but hereafter we simply refer to them as r and i bands.}
    \label{fig:filters}
\end{figure}

Here we make use of the full-depth/coverage data set of 
the Subaru SSP NB and BB (i.e., offband) images 
provided in the S21A internal release together with the newly processed CHORUS NB images, and select LAEs with the 
full data set at the completion of the survey. 
The information about NB filters used in this work is presented in Table \ref{tab:nb}, and their transmission curves are presented in Figure \ref{fig:filters}.
These LAEs are useful for statistical studies on
galaxy formation and cosmic reionization (H. Umeda et al. in preparation)
and are fed to the forthcoming massive spectroscopic program
of the Subaru Prime Focus Spectrograph \citep[PFS;][]{Takada2014} survey,
which complements the ongoing LAE spectroscopic program
of Hobby-Eberly Telescope Dark Energy Experiment \citep[HETDEX;][]{Gebhardt2021} in terms of depths, areas, and
redshifts \citep[cf.][]{Zhang2021c}.
%
%
%
%
%
%

Throughout this paper, we use the AB magnitude system and assume a cosmology with $\Omega_\mathrm{m} = 0.3$, $\Omega_\mathrm{\Lambda} = 0.7$, and $H_0 = 70 \mathrm{~km~s^{-1}~Mpc^{-1}}$,
unless otherwise noted.

\begin{table}[]
\caption{HSC Narrowband Filters Used in This Work}
\label{tab:nb}
\begin{tabular}{lccC}
\hline \hline
Filter & $\lambda_\mathrm{c}$ & FWHM & z_\mathrm{Ly\alpha} \\
& (\AA) & (\AA) &  \\ \hline
NB387 & 3862 & 55 & 2.18\pm0.023 \\ 
NB527 & 5260 & 79 & 3.33\pm0.032 \\
NB718 & 7170 & 111 & 4.90\pm0.046 \\
NB816 & 8177 & 113 & 5.72\pm0.046 \\
NB921 & 9215 & 135 & 6.58\pm0.056 \\
NB973 & 9713 & 112 & 6.99\pm0.046 \\
NB1010 & 10097 & 91 & 7.30\pm0.037 \\ \hline
\end{tabular}
\tablecomments{$\lambda_\mathrm{c}$: the central wavelengths of each filter; FWHM: the FWHMs of transmission curves (see Figure \ref{fig:filters}); $z_\mathrm{Ly\alpha}$: the redshift ranges of the Ly$\alpha$ emission lines that can be probed by each filter. }
\end{table}

\section{Data and LAE Selection} \label{sec:datasel}
\subsection{Survey Data and Source Detection}\label{subsec:datadet}
We combine the data products from the HSC-SSP S21A internal release with data obtained from the CHORUS project \citep{Inoue2020}. 
HSC-SSP is a three-tiered imaging survey consisting of the Wide, Deep (D), and UltraDeep (UD) layers.
In this study, we use data from the D and UD layers, which include observations with NB filters. 
In the D layer, three NB filters (NB387, NB816, and NB921) are used in addition to five BB filters \citep[grizy;][]{Kawanomoto2018}. In the UD layer, NB816, NB921, NB1010, and the five BBs are used, while the NB387 filter is not used. 
The D layer covers four separate fields: D-COSMOS, D-DEEP23, D-ELAISN1, and D-SXDS. 
The UD layer covers two areas in two fields in the D layer: UD-COSMOS and UD-SXDS. 
The CHORUS project is a deep HSC imaging survey that targets the UD-COSMOS field with four custom-made NB filters and one intermediate-band filter (NB387, NB527, NB718, NB973, and IB945), although the IB945 intermediate filter is not used in this work.

We use the HSC-SSP coadded images from the (currently) internal S21A release, which are reduced with hscPipe version 8, except for the NB387 images for the COSMOS field.
For the NB images taken in the CHORUS project, we download raw data and newly reduce them using the same hscPipe v8, because the public CHORUS images (CHORUS PDR1\footnote{\url{https://hsc-release.mtk.nao.ac.jp/doc/index.php/chorus/}}) are reduced with the previous version of hscPipe. 
Since the NB387 images of the UD-COSMOS field (the center of the four D-COSMOS pointings) are not observed by HSC-SSP \citep[and thus have a hole in the footprint; see][]{Aihara2022} but by CHORUS, we combine raw NB387 images from both surveys and newly produce the NB387 images of the COSMOS field without gaps of the observation footprints. 
We then correct a known systematic offset of $-0.45$ mag in a zero-point magnitude of our NB387 images \citep{Liang2021,Aihara2022}\footnote{See also the Known Problems page of HSC-SSP: \url{https://hsc-release.mtk.nao.ac.jp/doc/index.php/known-problems__pdr3/\#hsc-link-0}}. 
For the newly reduced COSMOS NB387 images, we correct an additional offset of $-0.13$ mag inferred from a comparison between the total magnitudes of stellar objects in the HSC-SSP PDR2 images and our data. 
The effective survey areas after masking (Section \ref{subsec:screening}) for each NB are listed in Table \ref{tab:surveyarea}.
The HSC images are processed in $1.7\times1.7\deg^2$ square regions called ``tracts'' in the current HSC pipeline. Tracts are predefined as an isolatitude tessellation, and each tract is further divided into $9\times9$ ``patches'' with 4100 or 4200 pixels on a side ($\sim11.5\arcmin$). 
Adjacent tracts have an overlap of one patch, and adjacent patches have an overlap of 100 pixels (17\arcsec) on their edges.
Our source detection and object selection are conducted on these patch images. 
The catalogs of the individual patches are combined after the selection (Section \ref{subsec:criteria}) and masking (Section \ref{subsec:screening}).

First, we measure the point-spread function (PSF) size from the FWHMs of point sources in all coadded images patch by patch, and we smooth the images with a Gaussian kernel to match the FWHMs to the largest PSF image (1{\mbox{$.\!\!\arcsec$}}0) to enable reliable measurement of source colors using aperture photometry. 
Next, we run SExtractor \citep{Bertin1996} in the dual image mode on the PSF-matched images using each NB as a detection band.
The important parameters of SExtractor are set to the following values: DETECT\_MINAREA$=5$, DETECT\_THRESH$=2.0$, and BACK\_SIZE$=64$. 
We use 2{\mbox{$.\!\!\arcsec$}}0-diameter aperture magnitudes to measure the colors of detected sources.

\begin{table*}[]
\caption{Effective survey area}
\label{tab:surveyarea}
\begin{tabular}{lccccccc}
\hline\hline
Field & NB387 & NB527 & NB718 & NB816 & NB921 & NB973 & NB1010 \\
 & (deg$^2$) & (deg$^2$) & (deg$^2$) & (deg$^2$) & (deg$^2$) & (deg$^2$) & (deg$^2$)  \\ \hline
 COSMOS & 7.14 & 1.79 & 1.76 & 6.66 & 6.70 & 1.72 & 1.76 \\ 
 DEEP2-3 & 6.01 & \nodata & \nodata & 5.68 & 5.70 & \nodata & \nodata \\
 ELAIS-N1 & 6.24 & \nodata & \nodata & 5.62 & 5.63 & \nodata & \nodata \\
 SXDS & 4.55 & \nodata & \nodata & 5.93 & 5.96 & \nodata & 1.76 \\\hline
Total & 23.94 & 1.79 & 1.76 & 23.89 & 23.99 & 1.72 & 3.52 \\\hline
\end{tabular}
\tablecomments{NB816 and NB921 cover the entire D and UD layers of the HSC-SSP survey. NB387 covers those fields except for the UD-SXDS field when combined with the CHORUS survey data. NB1010 covers the UD-COSMOS and UD-SXDS fields. NB527, NB718, and NB973 filters from the CHORUS survey cover the UD-COSMOS field.}
\end{table*}

Since our survey area is large and HSC-SSP has multiple layers with different depths, the limiting magnitudes within each survey field vary.
Adopting a single representative value as a limiting magnitude over the fields, as often done in previous studies, would result in nonuniform detections with more sources detected in low-signal-to-noise ratio (S/N) regions (e.g., near the edge
of the survey fields) with a high rate of spurious sources.
To address this issue, we measure the limiting magnitudes of all bands on a patch-by-patch basis. 
After masking low-S/N regions near the edges of the survey fields, 2{\mbox{$.\!\!\arcsec$}}0-aperture magnitudes of random points in each patch are measured. The number of random points in each patch is generally 10,000 or a smaller value depending on the area of masked regions. 
Their standard deviations $\sigma$ are obtained from the distribution of the aperture counts by fitting a Gaussian and used as $1\sigma$ 2{\mbox{$.\!\!\arcsec$}}0-aperture limiting magnitudes. 
We show the resulting $5\sigma$ limiting magnitude maps for all filters in Figures \ref{fig:limmag0387}--\ref{fig:limmagy} in Appendix \ref{sec:ap:limmag}. 
Galactic extinction correction for all sources is also performed on a patch-by-patch basis with $A_V$ values of patch centers from the \citet{Schlafly2011} calibration 
and assuming the extinction curve from \citet{Fitzpatrick1999} with $R_V=3.1$. 
We select LAE candidates based on these detection catalogs and limiting magnitude maps. 


\subsection{Photometric Selection Criteria of LAEs}\label{subsec:criteria}
We use the following photometric criteria to select LAE candidates consistently with previous studies \citep{Shibuya2018laelab,Itoh2018,Zhang2020b,Ono2021,Goto2021}:
\begin{eqnarray}
\begin{split}
\mathrm{NB387} < \mathrm{NB387_{6\sigma}}~\mathrm{AND} \\
\mathrm{g} - \mathrm{NB387} > \mathrm{max}(0.2, (\mathrm{g}-\mathrm{NB387})_\mathrm{3\sigma})~\mathrm{AND} \\
\mathrm{r} - \mathrm{NB387} > 0.3~\mathrm{AND} \\
\mathrm{i} - \mathrm{NB387} > 0.4~\mathrm{AND} \\
\mathrm{z} - \mathrm{NB387} > 0.5
\end{split}
\end{eqnarray}
for $z=2.2$ LAEs, 
\begin{eqnarray}
\begin{split}
\mathrm{NB527}  <  \mathrm{NB527_{5\sigma}}~\mathrm{AND} \\
\mathrm{g} - \mathrm{NB527} > \mathrm{max}(0.9, (\mathrm{g}-\mathrm{NB527})_\mathrm{3\sigma})~\mathrm{AND} \\
\mathrm{r} - \mathrm{NB527} > 0.3~\mathrm{AND} \\
\mathrm{i} - \mathrm{NB527} > 0.4~\mathrm{AND} \\
\mathrm{z} - \mathrm{NB527} > 0.5
\end{split}
\end{eqnarray}
for $z=3.3$ LAEs, 
\begin{eqnarray}
\begin{split}
\mathrm{g} > \mathrm{g_{2\sigma}}~\mathrm{AND}~\mathrm{NB718} < \mathrm{NB718_{5\sigma}}~\mathrm{AND} \\
\mathrm{ri} - \mathrm{NB718} > \mathrm{max}(0.7, (\mathrm{ri}-\mathrm{NB718})_\mathrm{3\sigma}) ~\mathrm{AND} \\
\mathrm{r} - \mathrm{i} > 0.8
\label{eq:z49}
\end{split}
\end{eqnarray}
for $z=4.9$ LAEs, where ri is calculated by the linear combination of the fluxes in the r band, $f_\mathrm{r}$, and those of the i band, $f_\mathrm{i}$, as $f_\mathrm{ri} = 0.3f_\mathrm{r} + 0.7f_\mathrm{i}$, 
\begin{eqnarray}
\begin{split}
\mathrm{g} > \mathrm{g_{2\sigma}}~\mathrm{AND} ~\mathrm{NB816} < \mathrm{NB816_{5\sigma}}~\mathrm{AND} \\
\mathrm{i} - \mathrm{NB816} > \mathrm{max}(1.2, (\mathrm{i}-\mathrm{NB816})_\mathrm{3\sigma})~\mathrm{AND} \\
((\mathrm{r} > \mathrm{r_{3\sigma}}) ~\mathrm{OR}~
(\mathrm{r} < \mathrm{r_{3\sigma}} ~\mathrm{AND}~
\mathrm{r} - \mathrm{i} > 1.0))
\end{split}
\end{eqnarray}
for $z=5.7$ LAEs, 
\begin{eqnarray}
\begin{split}
\mathrm{g} > \mathrm{g_{2\sigma}}~\mathrm{AND}~
\mathrm{r} > \mathrm{r_{2\sigma}}~\mathrm{AND} \\
\mathrm{NB921} < \mathrm{NB921_{5\sigma}}~\mathrm{AND} \\
\mathrm{z} - \mathrm{NB921} > \mathrm{max}(1.0, (\mathrm{z}-\mathrm{NB921})_\mathrm{3\sigma})~\mathrm{AND} \\
((\mathrm{z} > \mathrm{z_{3\sigma}}) ~\mathrm{OR}~
(\mathrm{z} < \mathrm{z_{3\sigma}} ~\mathrm{AND}~
\mathrm{i} - \mathrm{z} > 1.0))
\end{split}
\end{eqnarray}
for $z=6.6$ LAEs, 
\begin{eqnarray}
\begin{split}
\mathrm{g} > \mathrm{g_{2\sigma}}~\mathrm{AND}~ 
\mathrm{r} > \mathrm{r_{2\sigma}}~\mathrm{AND}~
\mathrm{i} > \mathrm{i_{2\sigma}}~\mathrm{AND} \\
\mathrm{NB973} < \mathrm{NB973_{5\sigma}}~\mathrm{AND} \\
((\mathrm{z} > \mathrm{z_{3\sigma}}) ~\mathrm{OR}~
(\mathrm{z} < \mathrm{z_{3\sigma}} ~\mathrm{AND}~
\mathrm{z} - \mathrm{y} > 2.0))~\mathrm{AND} \\
\mathrm{y} - \mathrm{NB973} > 0.7
\end{split}
\end{eqnarray}
for $z=7.0$ LAEs, 
and
\begin{eqnarray}
\begin{split}
\mathrm{g} > \mathrm{g_{2\sigma}}~\mathrm{AND}~
\mathrm{r} > \mathrm{r_{2\sigma}}~\mathrm{AND}~
\mathrm{i} > \mathrm{i_{2\sigma}}~\mathrm{AND}~
\mathrm{z} > \mathrm{z_{2\sigma}}~\mathrm{AND} \\
\mathrm{NB1010} < \mathrm{NB1010_{5\sigma}}~\mathrm{AND}~ \mathrm{y} - \mathrm{NB1010} > 1.9 \hspace{1.0cm}
\end{split}
\end{eqnarray}
for $z=7.3$ LAEs. 
When evaluating colors, if the BB magnitudes are fainter than $2\sigma$, we replace them with the $2\sigma$ limiting magnitudes. 
Equivalent width thresholds for each NB excess criterion roughly correspond to 10--20 \AA, depending on redshift (see Figures 2 and 3 of \citealt{Ono2021} and Figure 4 of \citealt{Goto2021}). Some faint objects are additionally affected by another criterion regarding the significance of NB excess. Users who intend to compare different redshift samples or select LAEs with a consistent equivalent width threshold across the sample should take these factors into account.
We remove unrealistically compact sources ($\mathrm{FWHM}<5$ pix) with measured FWHMs well below the sequence made by real point sources ($\mathrm{FWHM}=6$ pix) in an FWHM versus NB magnitude plot. 
For NB387 LAEs, we exclude sources that are very bright ($\mathrm{NB387}<18$ mag) and saturated in any of the NB or BB filters. In addition, we set the threshold for detection in NB387 to $6\sigma$ instead of $5\sigma$. This is to avoid detecting spurious sources in high noise regions; the detected source number as a function of significance of NB387 2{\mbox{$.\!\!\arcsec$}}0-aperture magnitude grows much faster below 6$\sigma$ due to the noise. 
The numbers of photometric LAE candidates at this step and those of the following steps (specified in Section \ref{subsec:screening}) are listed in Table \ref{tab:laenumber}.

\subsection{Screening}\label{subsec:screening}
The photometric candidates still contain a huge number of spurious sources, particularly near the edge of our survey field. 
We make masks to efficiently exclude spurious sources before visual inspection. 
The masks are placed in problematic regions of the images, where detected source density is higher than good regions. Specifically, low-S/N regions (near the survey field edge or due to pipeline failures), bright star halos, blooming features, channel-stop features (only in the y, NB973, and NB1010 filters), optical ghosts, and satellite trails are masked. 
Our masks for the four SSP NB filters are made based on the official Bright Star Masks\footnote{\url{https://hsc-release.mtk.nao.ac.jp/doc/index.php/bright-star-masks__pdr3/}}, with additional regions added to remove artifacts that are not included in the official masks, and unnecessary official masks deleted or reduced in size. 
For CHORUS NBs, we define masks by ourselves paying particular attention to the high (spurious) source density regions. 
At this stage, we combine the catalogs of all patches for the four separate fields (COSMOS, DEEP2-3, ELAIS-N1, SXDS, with no distinction of D and UD), 
and remove sources that appear twice or more due to patch and tract overlap with a matching radius of 0{\mbox{$.\!\!\arcsec$}}336 (2 pixels). 
For a pair of objects from different patches with separation $<$0{\mbox{$.\!\!\arcsec$}}336, we exclude sources with higher MAGERR\_APER measured by SExtractor from the catalogs.
This process reduces the number of sources by $\sim25$\%.
In \citet{Ono2021}, patch-by-patch limiting magnitude measurements and masking of problematic regions are not performed, resulting in a large number of photometric candidates and the use of machine learning techniques. 
As Table \ref{tab:laenumber} shows, we can significantly reduce the number of candidates for visual inspection by appropriately setting the limiting magnitudes and using masks ($90-99.7$\%). 
Since the sample size of our candidates at this stage is already comparable to that of \citet{Ono2021} after the machine learning selection, we directly proceed to visual inspection. 

Our visual inspection is conducted by four individuals (S.K., Y.L., H.U., A.M.).
The candidates are classified as either secure, ambiguous, or spurious sources. 
Objects are labeled as ``spurious'' when they are identified as cosmic rays, sources with a sharp edge (probably due to some pipeline failures), sources detected in problematic regions remaining after masking, or any other signals that seem to be artificial. 
Sources that appear to be associated with resolved nearby galaxies are classified as spurious. 
Additionally, if objects are sufficiently bright (and hence potentially astrophysically interesting) but could be moving or transient objects that are identified from their shapes or unphysically high NB excesses, we inspect warp images (see the next paragraph) and label them as spurious if their nonstatic nature is confirmed. 
Objects are labeled as ``ambiguous'' if they are very diffuse or hard to distinguish from noise and ghosts.
Consistencies between the classifications by each individual are checked with NB816 LAE candidates in the COSMOS field. 
We achieve a $>96\%$ consistency, with the inconsistent cases mostly being faint ambiguous sources or located at problematic regions. 
The fractions of survived sources are, from the NB387 to NB1010 sample, 55\%, 76\%, 92\%, 50\%, 52\%, 22\%, and 0.90\% (Table \ref{tab:laenumber}), although five candidates in the NB1010 sample are very tentative (see discussion in Section \ref{subsec:nb1010}).
This variation in the survived fraction is caused by the balance between real and spurious source density. 
While the intrinsic rarity of $z\gtrsim7$ LAEs makes the fraction quite low, the selection criteria that necessitate detections in more than two bands effectively reduce the number of spurious sources because the probability of random fluctuations causing such detections is extremely low (e.g., the criterion $\mathrm{r}-\mathrm{i}>0.8$ in Equation \ref{eq:z49} implicitly requires the i band to be brighter than $\mathrm{r_{2\sigma}}-0.8$).

Since bright LAEs are rare, they are more susceptible to  contamination by artificial signals and nonstatic sources, such as moving objects, than numerous fainter sources. 
To further screen out any remaining contaminants in bright candidates, we visually inspect their NB photometry in multiepoch images taken over a period of up to 7 yr for the HSC-SSP NBs. 
The numbers of objects inspected in each NB are 379, 318, 110, and 5 for NB387 (all candidates with $\mathrm{NB387}<22.5$ 2{\mbox{$.\!\!\arcsec$}}0-aperture mag), NB816 ($<24.5$ 2{\mbox{$.\!\!\arcsec$}}0-aperture mag), NB921 ($<24.5$ 2{\mbox{$.\!\!\arcsec$}}0-aperture mag), and NB1010 (all), respectively.
Objects that do not show consistent brightness and/or peak locations in multiepoch images (except for cases with pipeline failures or when the source falls on the CCD gaps or any other unusable regions; such shots are mostly clipped before coadding) are excluded from our catalogs. 
The numbers of excluded sources are 1, 20, 28, and 0 for NB387, NB816, NB921, and NB1010, respectively. 
We also try to identify contaminants with warp images for the CHORUS NBs, although the time spans of imaging observations of the CHORUS NBs are shorter than HSC-SSP's duration (3 days for NB718 and NB973, and 3 month for NB527, but most of the NB527 images are taken within 3 days, with only 2 shots taken 3 months before those observations being available). 
We check 212, 101, 23 objects for NB527 (all candidates with $\mathrm{NB527}<24.5$ 2{\mbox{$.\!\!\arcsec$}}0-aperture mag), NB718 ($\mathrm{NB718}<25.0$ 2{\mbox{$.\!\!\arcsec$}}0-aperture mag), and NB973 (all), respectively. 
We exclude 1, 0, and 5 objects in the NB527, NB718, and NB973 samples, respectively. 
This completes our cataloging process. 


\begin{table*}
    \caption{Numbers of Candidates Remaining after Each Step and Final Samples}
\label{tab:laenumber}
\begin{tabular}{lrrrrrrr}
\hline
Filter & NB387 & NB527 & NB718 & NB816 & NB921 & NB973 & NB1010 \\ \hline\hline
Photometric selection & 176843 & 28346 & 2577 & 208675 & 174137 & 41232 & 67409 \\ 
Masking & 16552 & 7153 & 934 & 16156 & 5192 & 123 & 680 \\
Overlap Rejection & 12599 & 6096 & 789 & 12325 & 3989 & 106 & 556 \\
Visual Inspection & 6996 & 4642 & 726 & 6144 & 2086 & 23 & 5 \\ \hline
Final sample & 6995 & 4641 & 726 & 6124 & 2058 & 18 & \tablenotemark{*}5 \\ \hline
COSMOS & 3160 & 4641 & 726 & 1625 & 547 & 18 & 3 \\
DEEP2-3  & 1093 & \nodata & \nodata & 1551 & 494 & \nodata & \nodata \\
ELAIS-N1  & 1767 & \nodata & \nodata & 1354 & 468 & \nodata & \nodata \\
SXDS  & 975 & \nodata & \nodata & 1594 & 549 & \nodata & 2 \\ \hline
\end{tabular}
\tablenotetext{*}{Tentative candidates (see Section \ref{subsec:nb1010})}
\end{table*}

\section{Results and Discussion} \label{sec:results}
\subsection{Sky Distributions and Number Counts}\label{subsec:mapnc}

\begin{figure*}
\plotone{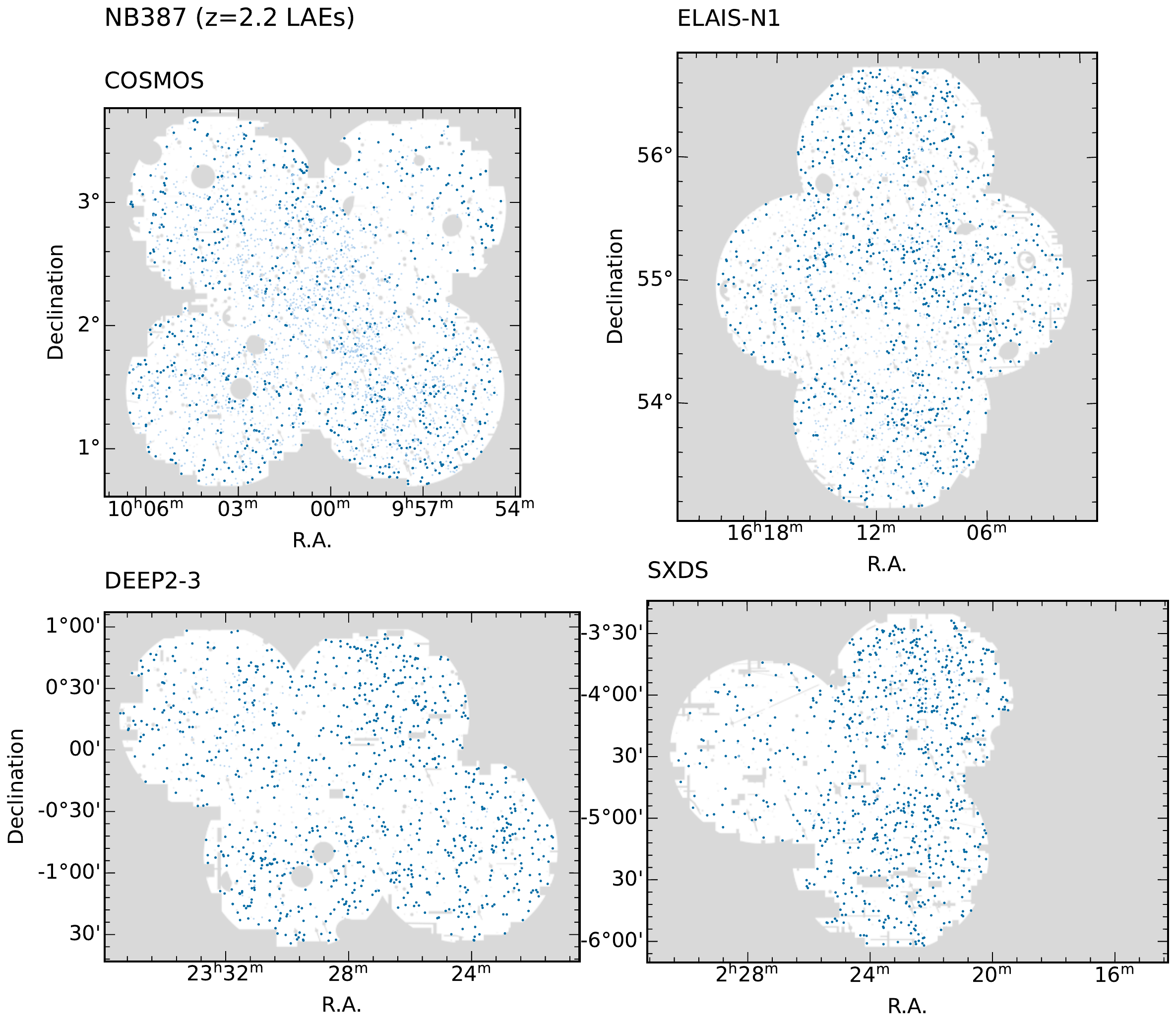}
\caption{Sky distribution of NB387 LAEs ($z=2.2$). Thick (thin) blue dots indicate the locations of objects brighter (fainter) than $\mathrm{NB387}=24.0$ mag, respectively. North is up, and east to the left. Masked regions are shown in gray.}
\label{fig:skydis387}
\end{figure*}

\begin{figure*}
\plotone{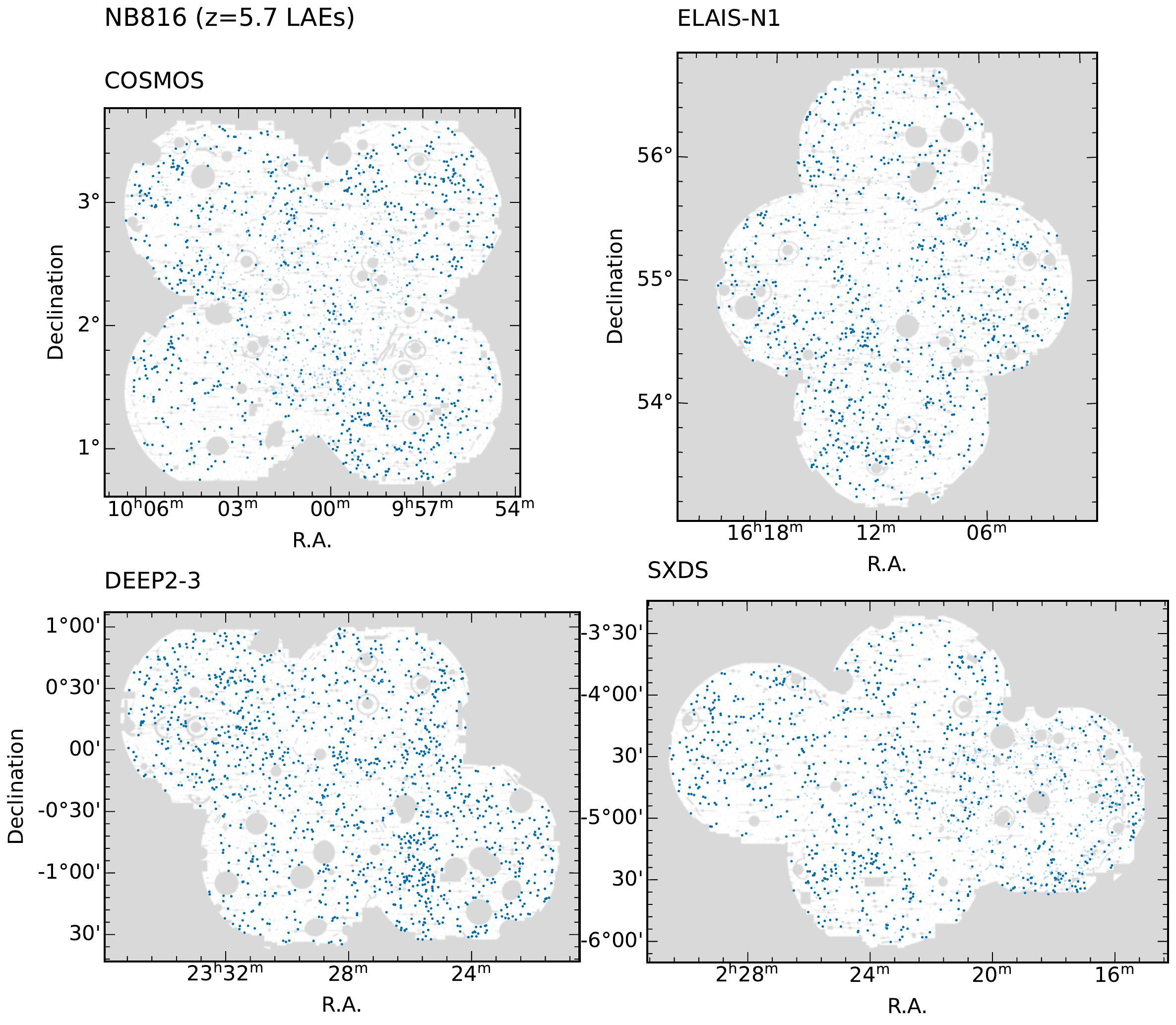}
\caption{Sky distribution of NB816 LAEs ($z=5.7$). Thick (thin) blue dots indicate the locations of objects brighter (fainter) than $\mathrm{NB816}=25.5$ mag, respectively. North is up, and east to the left. Masked regions are shown in gray.}
\label{fig:skydis816}
\end{figure*}

\begin{figure*}
\plotone{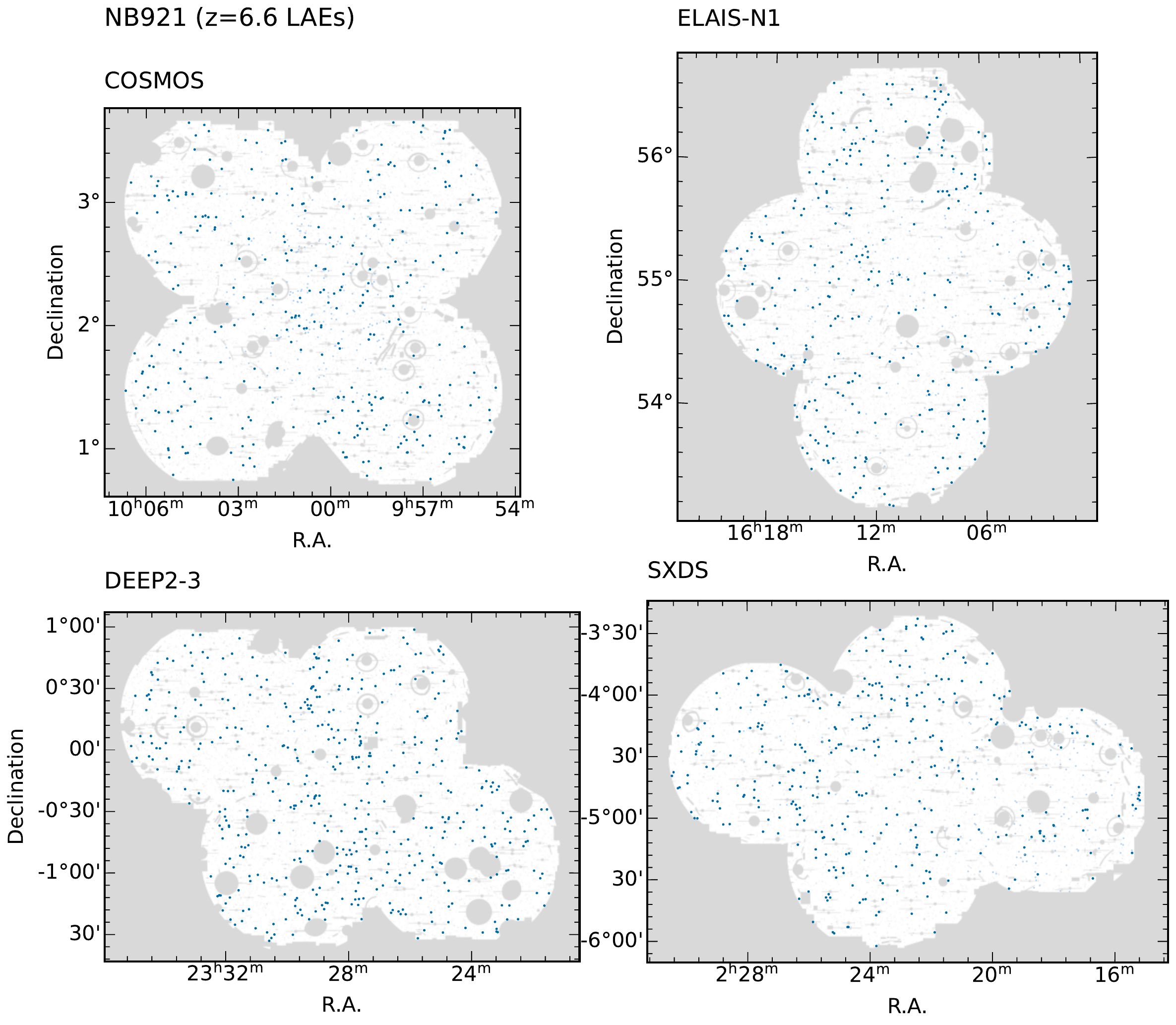}
\caption{Sky distribution of NB921 LAEs ($z=6.6$). Thick (thin) blue dots indicate the locations of objects brighter (fainter) than $\mathrm{NB921}=24$ mag, respectively. North is up, and east to the left. Masked regions are shown in gray.}
\label{fig:skydis921}
\end{figure*}

\begin{figure*}
\plotone{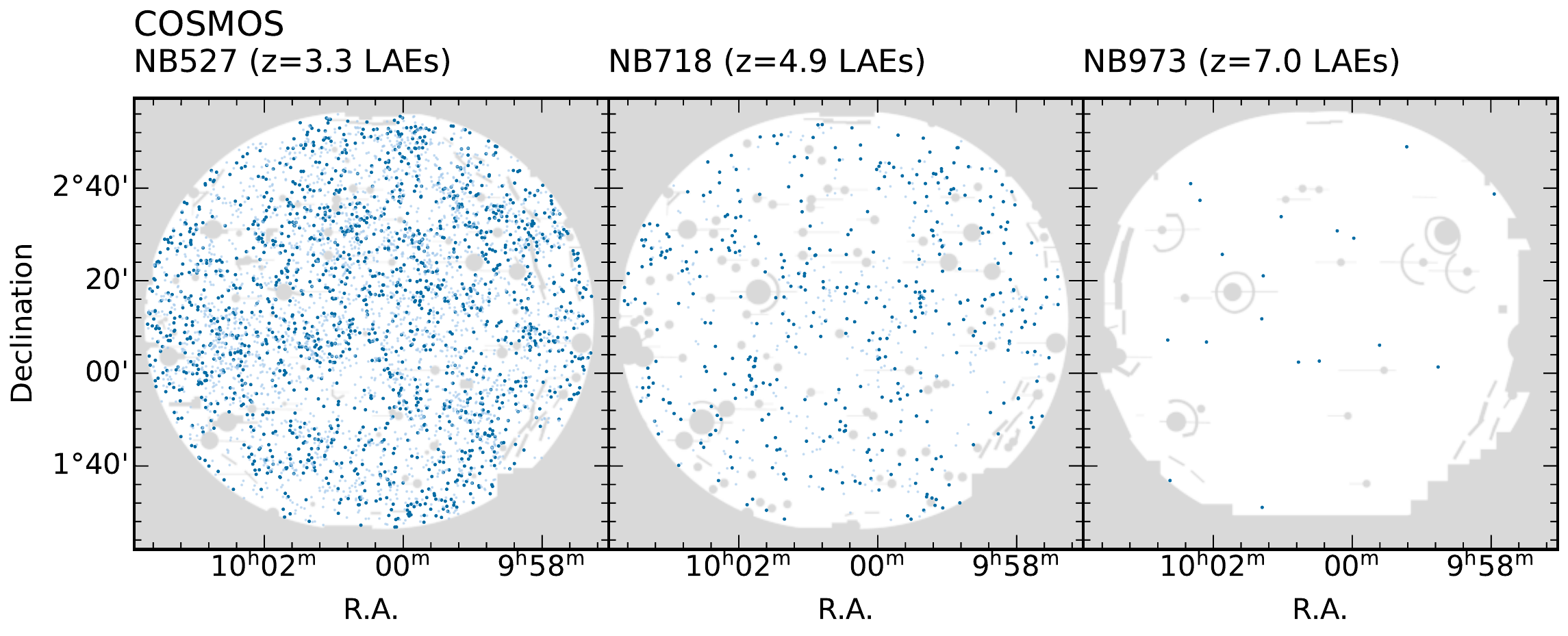}
\caption{From left to right, sky distribution of NB527, NB718, and NB973 LAEs ($z=3.3, 4.9$, and 7.0). Thick (thin) blue dots indicate the locations of objects brighter (fainter) than NB527 or $\mathrm{NB718}=25.75$ mag, while all sources are plotted with thick blue dots in the NB973 map. North is up, and east to the left. Masked regions are shown in gray.}
\label{fig:skydischorus}
\end{figure*}

The numbers of LAE candidates at $z=2.2$, 3.3, 4.9, 5.7, 6.6, 7.0, and 7.3 after screening are 6995, 4641, 726, 6124, 2058, 18, and 5, respectively. 
We also list the numbers of LAEs in each field in Table \ref{tab:laenumber}. 
The sky distributions of our secure LAEs are presented in Figures \ref{fig:skydis387}--\ref{fig:skydischorus}.
Sufficiently bright candidates ($<24.0$ mag for the NB387 sample, $<25.75$ mag for the NB527 and NB718 samples, and $<25.5$ mag for the NB816 and NB921 samples) that can be homogeneously selected over the entire fields are plotted with thick blue dots, while faint candidates, which are especially more numerous in the UD layer, are plotted with thin blue dots for clarity.
There are no clear underdensities of real sources or overdensities made of spurious sources near the edges of the HSC pointings, ensuring that our selection with spatially dependent detection limits (Figures \ref{fig:limmag0387}--\ref{fig:limmag1010}) and the screening processes specified in Section \ref{subsec:screening} successfully provide clean samples. 

\begin{figure}[ht!]
\plotone{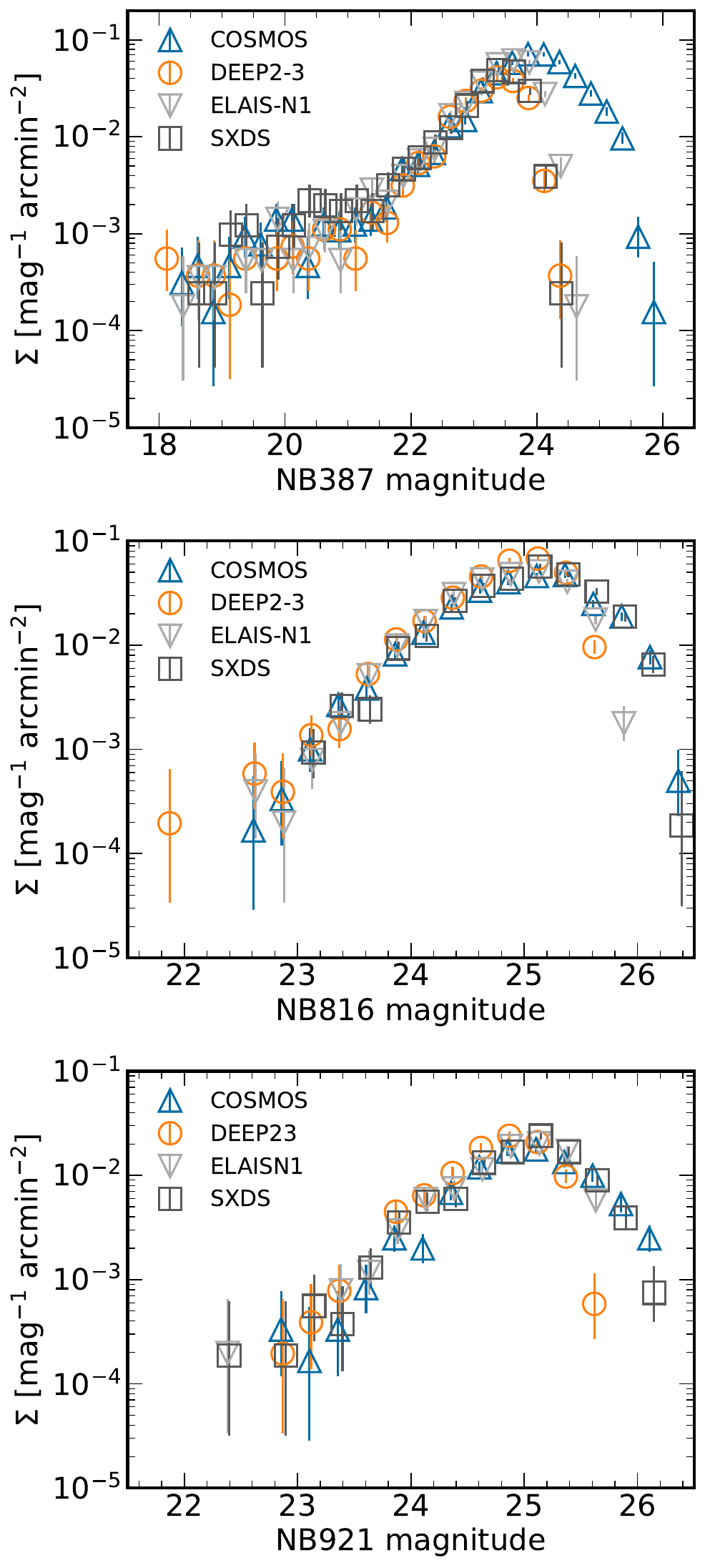}
\caption{
  Number counts of our LAEs in each subfield. From top to bottom, results for NB387, NB816, and NB921 are shown. Here we use total magnitude (MAG\_AUTO of SExtractor) for the x-axis value. Symbols for the same x-axis value are slightly offset for clarity.}
\label{fig:number_counts}
\end{figure}

\begin{figure}[ht!]
\includegraphics[width=0.45\textwidth]{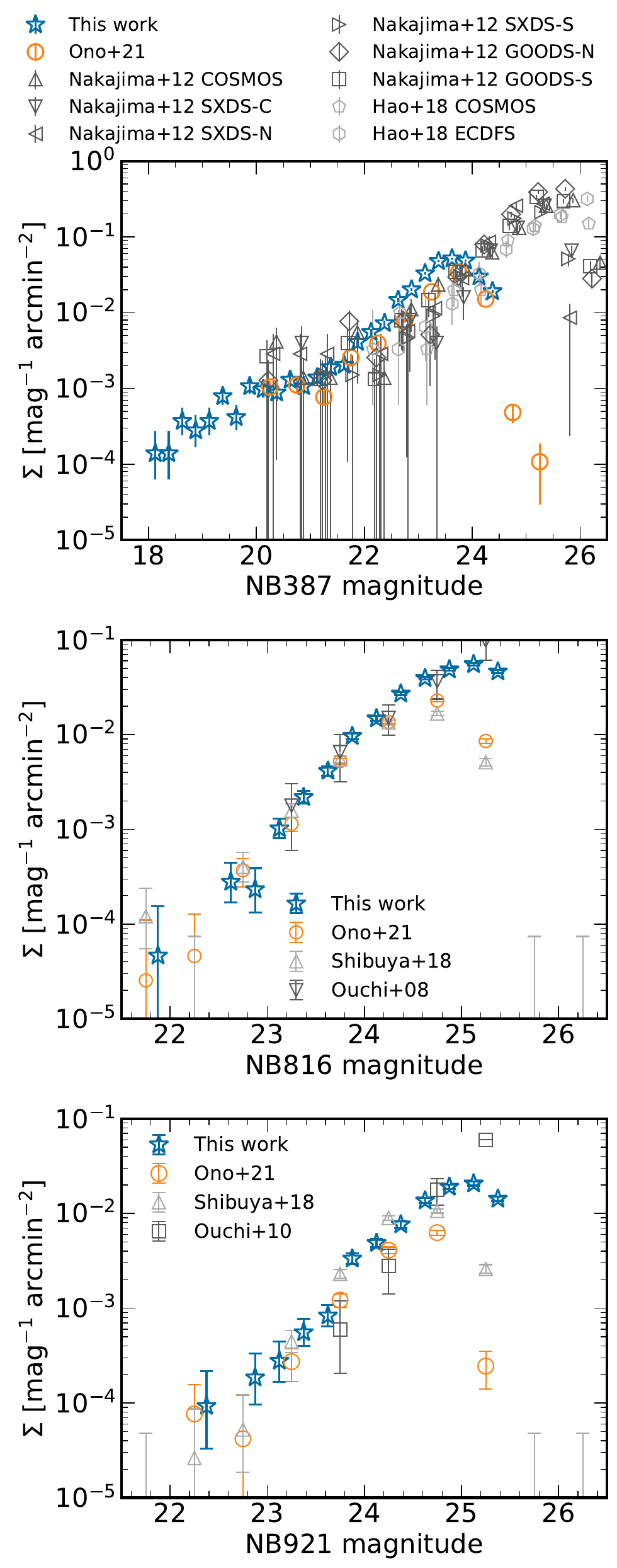}
\caption{
  Similar to Figure \ref{fig:number_counts}, but the mean number counts of our LAEs are compared with those of similar previous LAE surveys. From top to bottom, results for NB387, NB816, and NB921 are shown. Here we use total magnitude (MAG\_AUTO of SExtractor) for the x-axis value for a fair comparison. References for the NB387 results: \citet{Nakajima2012, Hao2018}; References for the NB816 results: \citet{Ouchi2008,Shibuya2018laelab,Ono2021}; References for the NB921 results: \citet{Ouchi2010,Shibuya2018laelab,Ono2021}. }
\label{fig:number_counts_previous}
\end{figure}

\begin{figure}[ht!]
\includegraphics[width=0.45\textwidth]{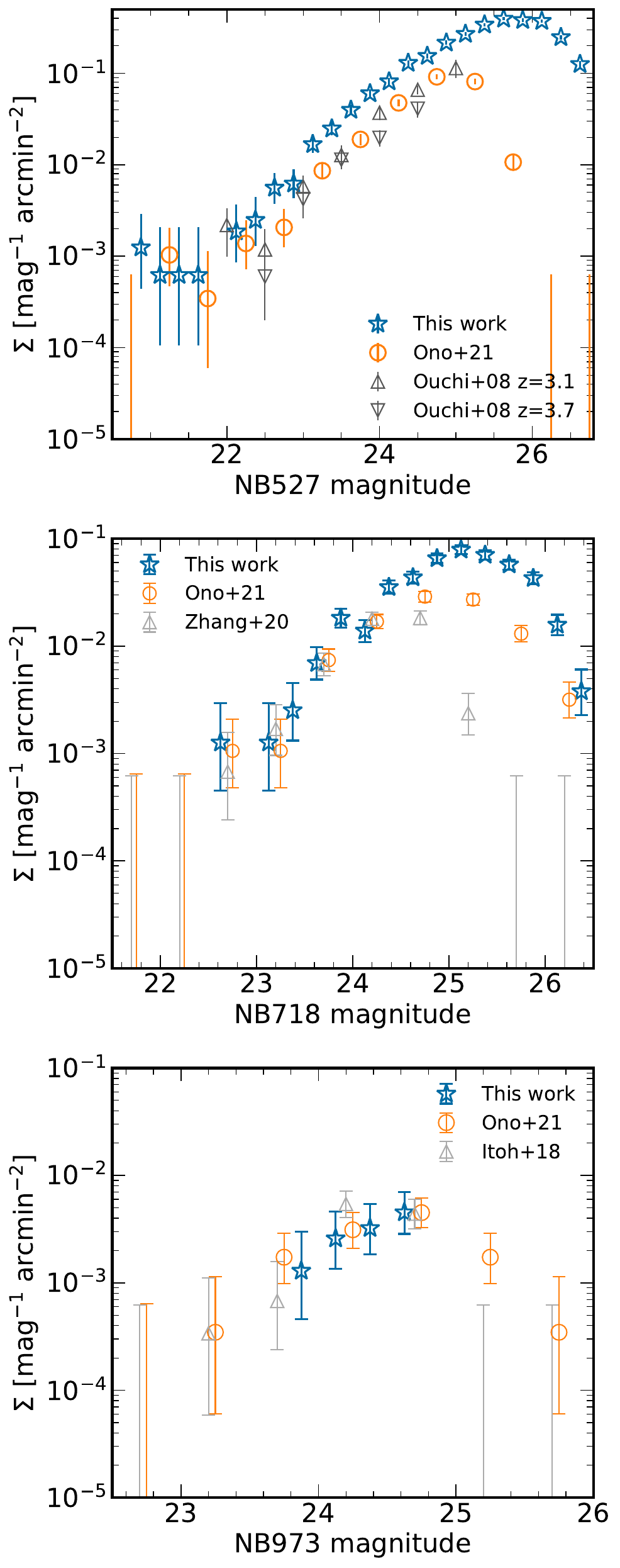}
\caption{
  Similar to Figure \ref{fig:number_counts_previous}, but showing the result for the CHORUS NBs. From top to bottom, results for NB527, NB718, and NB973 are shown. Here we use total magnitude (MAG\_AUTO of SExtractor) for the x-axis value for a fair comparison. References for the NB527 results: \citet{Ouchi2008,Ono2021}; References for the NB718 results: \citet{Zhang2020b,Ono2021}; References for the NB973 results: \citet{Itoh2018,Ono2021}. }
\label{fig:number_counts_chorus}
\end{figure}

Figure \ref{fig:number_counts} presents the surface number density, or number counts, of our LAEs in the COSMOS, DEEP2-3, ELAIS-N1, and SXDS subfields for the three SSP NB filters. 
The number counts are in good agreement among the subfields, indicating that the LAEs are uniformly sampled within
the galaxy clustering signals that are quantified 
in H. Umeda et al. (in preparation). 
We compare the average number counts over our survey fields with those from previous studies in Figure \ref{fig:number_counts_previous} (SSP NBs) and Figure \ref{fig:number_counts_chorus} (CHORUS NBs).
Our number counts are largely consistent with those from previous studies, with differences mostly due to variations in selection criteria, survey depths, and cosmic variance.
For instance, our number count for NB387 shows a higher value than previous studies at the fainter end, which could be explained by the difference in selection criteria. Specifically, \citet{Nakajima2012} and \citet{Hao2018} use the u and B bands to constrain the Lyman break of LAEs and remove some low-redshift contaminants, whereas we and \citet{Ono2021} do not. 
Our number counts for NB527 show a factor of $\sim$2 excess compared to others, but this discrepancy can be attributed to differences in selection criteria, such as the use of the B band and stricter NB excess criteria in \citealt{Ouchi2008}. Indeed, the excess is mitigated if we adopt a similarly strict NB excess criteria ($\mathrm{g}-\mathrm{NB527}>1.3$) as in \citet{Ouchi2008}. 
On the other hand, \citet{Ono2021} use selection criteria that were almost identical to ours and the data from the HSC-SSP PDR2 and the CHORUS PDR1. We notice that our bright $z=3.3$ LAEs (where photometric errors are negligible) are also listed in the original detection catalog of \citet{Ono2021} and meet the selection criteria even in their S18A photometry, but $\sim50\%$ of them are not included in their final LAE catalog. This indicates that their machine learning selection method misses a considerable fraction of the good candidates, highlighting the limitations of the technique. 
One possible reason for this could be that their training data sets do not include major contaminants such as stellar halos and ghosts or atypical real LAEs such as Ly$\alpha$ blobs (LABs), leading to lower purity and completeness. 
To improve the performance of machine learning techniques, a prior understanding of the characteristics of typical contaminants and the objects of interest is necessary.

\subsection{Reliability of the NB1010 ($z=7.3$) Candidates}\label{subsec:nb1010}
We identify five tentative candidates at $z=7.3$ with NB1010 detection less than $<5.4\sigma$ and no detection above $2\sigma$ in the five BBs, including the y band ($<1.1\sigma$ for all candidates). 
However, given their diffuse appearance, caution is warranted. 
To assess the significance of these sources, we invert the NB1010 images and attempt to identify LAE candidates using the same selection criteria and screening processes. This leads to the identification of the comparable number of candidates, which suggests that the detection of the five sources could be explained by background fluctuations. 
If they are real, although the result of this test and nondetection of lower-equivalent width LAEs which should be more abundant suggest otherwise, their nondetection in the y band (all below $1.1\sigma$, which is statistically consistent with zero flux in the y band) suggests very high lower limits on their Ly$\alpha$ equivalent widths \citep[see Figure 4 of][]{Goto2021}. 
When their y-band flux is replaced with $2\sigma$ upper limits, $\mathrm{y_{2\sigma}-NB1010}$ values range from 2.6--2.9, corresponding to lower limits on Ly$\alpha$ equivalent widths of 50--300 \AA, even if they are precisely at $z=7.30$. These values are close to the physically possible upper limit of 3.1 (corresponding to infinite equivalent width). 
LAEs with Ly$\alpha$ equivalent width $>300$ \AA\ are exceptionally rare \citep{Hashimoto2017}, but the recent JWST discovery of a $z=7.3$ LAE with Ly$\alpha$ equivalent width of $\sim400$ \AA\ by \citet{Saxena2023} and our tentative detection, if real, imply that such extreme LAEs could be common during the epoch of reionization. 
In either case, the constraint on the number density of $z=7.3$ LAEs from this study is consistent within $1\sigma$ with \citet{Shibuya2012} and \citet{Goto2021}. Assuming that the NB1010 flux is solely due to Ly$\alpha$ emission, our results roughly correspond to the cumulative number density of $z=7.3$ LAEs with $\log L_\mathrm{Ly\alpha}\gtrsim43.1$ of $<1.1\times10^{-6}~\mathrm{Mpc^{-3}}$ (zero detection case) and $1.6-4.8\times10^{-6}~\mathrm{Mpc^{-3}}$ 
\citep[five detection case; see Figure 6 of][]{Goto2021}.

\subsection{Matching with Spectroscopic Catalogs}\label{subsec:spec}
To further ensure the validity of our catalog, we crossmatch our LAEs with galaxies and AGNs in publicly available spectroscopic catalogs \citep{Ouchi2008,Saito2008,Lilly2009,Ouchi2010,Willott2010,Coil2011a,Mallery2012,Masters2012,Newman2013,LeFevre2013,McGreer2013,Willott2013,Shibuya2014,Kriek2015,Liske2015,Sobral2015,Banados2016,Hu2016,Momcheva2016,Toshikawa2016,Wang2016,Hu2017a,Jiang2017,Masters2017,Suzuki2017,Tasca2017,Yang2017d,Harikane2018,Hasinger2018,Jiang2018,Lee2018,Pentericci2018,Scodeggio2018,Shibuya2018laelab,Sobral2018,Calvi2019,Harikane2019,Higuchi2019,Kashino2019fc,Masters2019,Lyke2020,Onodera2020,Zhang2020b,Ning2020a,Ono2021,Ning2022a} and our past spectroscopic observations with Magellan/IMACS \citep[PI: M. Ouchi; M. Ouchi et al. in preparation;]{Ono2018,Harikane2019}, using a matching radius of 1.0\arcsec.
In our photometric catalogs of LAEs, we identify 50, 8, 13, 181, 36, and 1 spectroscopically
confirmed LAEs at $z=2.2$, 3.3, 4.9, 5.6, 6.6, and 7.0, respectively. 
The properties of these sources, including their spectroscopic redshifts, are listed in Table \ref{tab:spec} in Appendix \ref{sec:ap:specmatch}. 
Figure \ref{fig:zhist} shows the redshift distributions of the spectroscopically identified sources, along with filter transmission curves displayed with thick-colored curves. 
The peaks of the distributions show slight offsets toward the lower-redshift side compared to the filter transmission curves. This well-known behavior is caused by the asymmetric Ly$\alpha$ line profile and Lyman break, which lead the Ly$\alpha$ red wing and UV continuum redward of the Lyman break of LAEs at lower redshifts to contribute more to the NB flux \citep{Ouchi2008, Kashikawa2011}. 
We also show the slightly shifted transmission curves, as in \citet{Ono2021}, with thin-colored curves that better match the redshift distributions. 
The amounts of the shifts are $-0.01, -0.02, -0.03, -0.02$, and $-0.01$ for the NB387, NB527, NB718, NB816, and NB921 filters, respectively.
This reasonable agreement between the redshift distributions and the transmission curves further assures the reliability of our selection.

We also found 53 low-redshift objects in the NB387 and NB527 samples, whose properties are listed in Table \ref{tab:lowzspec} in Appendix \ref{sec:ap:specmatch}. 
Most of these objects are CIV$\lambda$1549\AA\ emitters at $z\sim1.5$ and 2.4, as well as CIII]$\lambda$1909\AA\ emitters at $z\sim1.0$. 
This type of contamination is expected because our photometry in bands bluer than NB387 and NB527, and therefore the Lyman break of the $z=2.2$ and 3.3 sample, is not strongly constrained. 
These high-ionization line emitters that meet our criteria typically have relatively high equivalent widths ($\gtrsim30$\AA) and are thus considered to be AGNs \citep{Nakajima2018b}. In fact, most of the contaminants come from the Sloan Digital Sky Survey DR16 quasar catalog \citet{Lyke2020} or have broad line features.
The intrinsic number density of such sources should be much smaller than that of the target LAEs, and the survey volume for lower-redshift line emitters is smaller than that for LAEs at $z=2.2$ and 3.3 due to cosmological redshift. 
Contamination from [OII] emitters at $z=0.03$ is expected to be negligible for the same reason ($\sim0.2$\% of the volume for LAEs).
The contamination rate for $z=5.7$ and $z=6.6$ LAEs are estimated to be 14\% and 8\%, respectively, in \citet{Shibuya2018spec} based on spectroscopic follow-up observations of LAEs selected with almost the same selection criteria as ours. 
However, all the five bright contaminants reported in \citet{Shibuya2018spec} are successfully rejected in our catalog thanks to the deeper BB images, which provide strong constraints on their photometry below the Lyman break. 
Our LAEs at $z=5.7$ and 6.6 should have a much lower level of contamination, which would be acceptable for various statistical studies. 
This will be directly assessed in future massive spectroscopic follow-up observations in the planned PFS-SSP survey \citep[e.g., ][]{Greene2022}.

\begin{figure*}
\includegraphics[width=\textwidth]{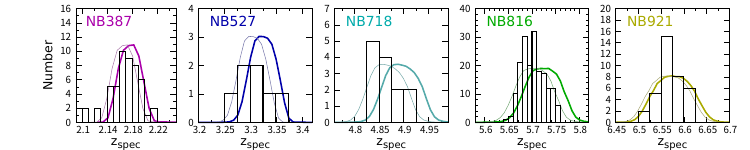}
\caption{
Redshift distribution of spectroscopically confirmed LAEs in our catalogs. Thick-colored curves show the normalized transmission curves of HSC NB filters (Figure \ref{fig:filters}) with wavelength converted to Ly$\alpha$ redshift, while thin-colored curves show the shifted ones.
}
\label{fig:zhist}
\end{figure*}


\subsection{Data Release and Example Science Cases}\label{subsec:dr}
Our LAE catalogs will be available online via 
\url{http://cos.icrr.u-tokyo.ac.jp/rush.html} after the next HSC-SSP official data release (Public Data Release 4). 
These LAE catalogs include the basic photometric properties with the coordinates together with ancillary information such as the limiting magnitude maps and the masks used in Section \ref{subsec:screening}.
Using these LAE catalogs, we will constrain the statistical properties of LAEs at various redshifts such as Ly$\alpha$ luminosity functions, clustering properties, and Ly$\alpha$ equivalent width distributions (\citealt{Konno2018, Ouchi2018, Shibuya2018laelab}, H. Umeda et al, in preparation) with unprecedentedly small statistical uncertainties. 
Ionized bubble size in the epoch of reionization will be, for the first time, constrained with enough sensitivity by cross-correlating the spatial distribution of LAEs spectroscopic follow-up observations conducted with the coming PFS-SSP survey and HI 21 cm emission observations obtained with the Square Kilometre Array \citep{Kubota2018, Kubota2020}. 
Deep imaging data of HSC-SSP and CHORUS can also be used to constrain diffuse nebula emission around LAEs and other galaxy populations \citep[e.g.][]{Kikuchihara2020, Kikuta2019, Kikuta2023}.

\section{Summary}
We construct massive Ly$\alpha$ emitter (LAE) catalogs at various redshifts using the latest deep images from the completed HSC-SSP survey \citep{Aihara2022} and the CHORUS survey \citep{Inoue2020}. 
The unique narrow-band filter sets of HSC allow us to probe LAEs at $z=2.2$, 3.3, 4.9, 5.7, 6.6, 7.0, and 7.3. 
CHORUS NB387 data are combined to fill the unobserved region at COSMOS in HSC-SSP NB387 images. 
After properly accounting for the spatial variation of limiting magnitudes, we initially select 699,219 photometric LAE candidates. 
Careful screening via masking of problematic regions and visual inspection of coadded and multiepoch images yields a total of 20,567 LAEs, consisting of 6995, 4641, 726, 6124, 2058, 18, and 5 LAEs at $z=2.2$, 3.3, 4.9, 5.7, 6.6, 7.0, and 7.3, respectively, although the $z=7.3$ candidates are tentative (Section \ref{subsec:nb1010}). 
The LAE number counts agree well with those from previous studies once the difference in selection and depth are taken into account (Section \ref{subsec:mapnc}), and our catalogs include 289 spectroscopically confirmed LAEs (Section \ref{subsec:spec}), and validate our selection processes. 
Our new catalogs will be made public on our project website (Section \ref{subsec:dr}). 

\begin{acknowledgments}
We thank the anonymous referee for careful reading and useful comments, and Ryohei Itoh for sharing his data and Ken Mawatari for his help with data analyses.
This work is supported by World Premier International Research Center Initiative (WPI Initiative), MEXT, Japan, as well as the joint research program of the Institute of Cosmic Ray Research (ICRR), the University of Tokyo. This study is supported by KAKENHI (19H00697, 20H00180, and 21H04467) Grant-in-Aid for Scientific Research (A) through the Japan Society for the Promotion of Science.  

The Hyper Suprime-Cam (HSC) collaboration includes the astronomical communities of Japan and Taiwan, and Princeton University.  The HSC instrumentation and software were developed by the National Astronomical Observatory of Japan (NAOJ), the Kavli Institute for the Physics and Mathematics of the Universe (Kavli IPMU), the University of Tokyo, the High Energy Accelerator Research Organization (KEK), the Academia Sinica Institute for Astronomy and Astrophysics in Taiwan (ASIAA), and Princeton University.  Funding was contributed by the FIRST program from the Japanese Cabinet Office, the Ministry of Education, Culture, Sports, Science and Technology (MEXT), the Japan Society for the Promotion of Science (JSPS), Japan Science and Technology Agency  (JST), the Toray Science  Foundation, NAOJ, Kavli IPMU, KEK, ASIAA, and Princeton University.

This work is based on data collected at the Subaru Telescope and retrieved from the HSC data archive system, which is operated by Subaru Telescope and Astronomy Data Center (ADC) at NAOJ. Data analysis was in part carried out with the cooperation of Center for Computational Astrophysics (CfCA) at NAOJ.  We are honored and grateful for the opportunity of observing the Universe from Maunakea, which has cultural, historical, and natural significance in Hawaii.

This paper makes use of software developed for Vera C. Rubin Observatory. We thank the Rubin Observatory for making their code available as free software at \url{http://pipelines.lsst.io/}. 
\end{acknowledgments}

%

\vspace{5mm}
\facilities{Subaru(HSC)}


\software{astropy \citep{Robitaille2013,Price-Whelan2018a}, 
          Source Extractor \citep{Bertin1996}
          }



\appendix

\section{Limiting Magnitude Maps for Narrowband and Broadband Filters}\label{sec:ap:limmag}
\begin{figure*}[]
    \plotone{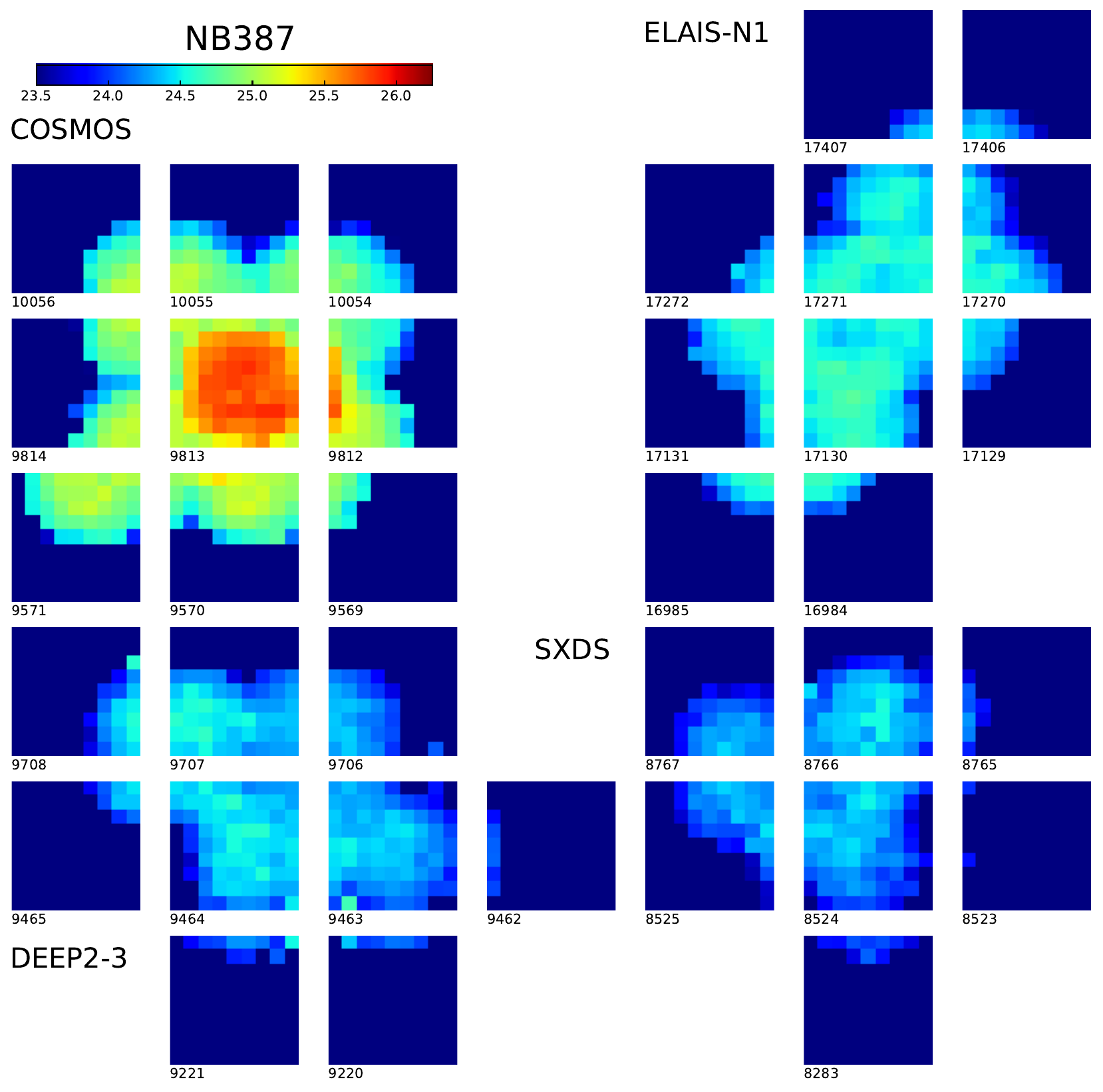}
    \caption{Limiting magnitude maps ($5\sigma$, 2{\mbox{$.\!\!\arcsec$}}0 aperture) for the NB387 images.}
    \label{fig:limmag0387}
\end{figure*}

\begin{figure*}[]
    \plotone{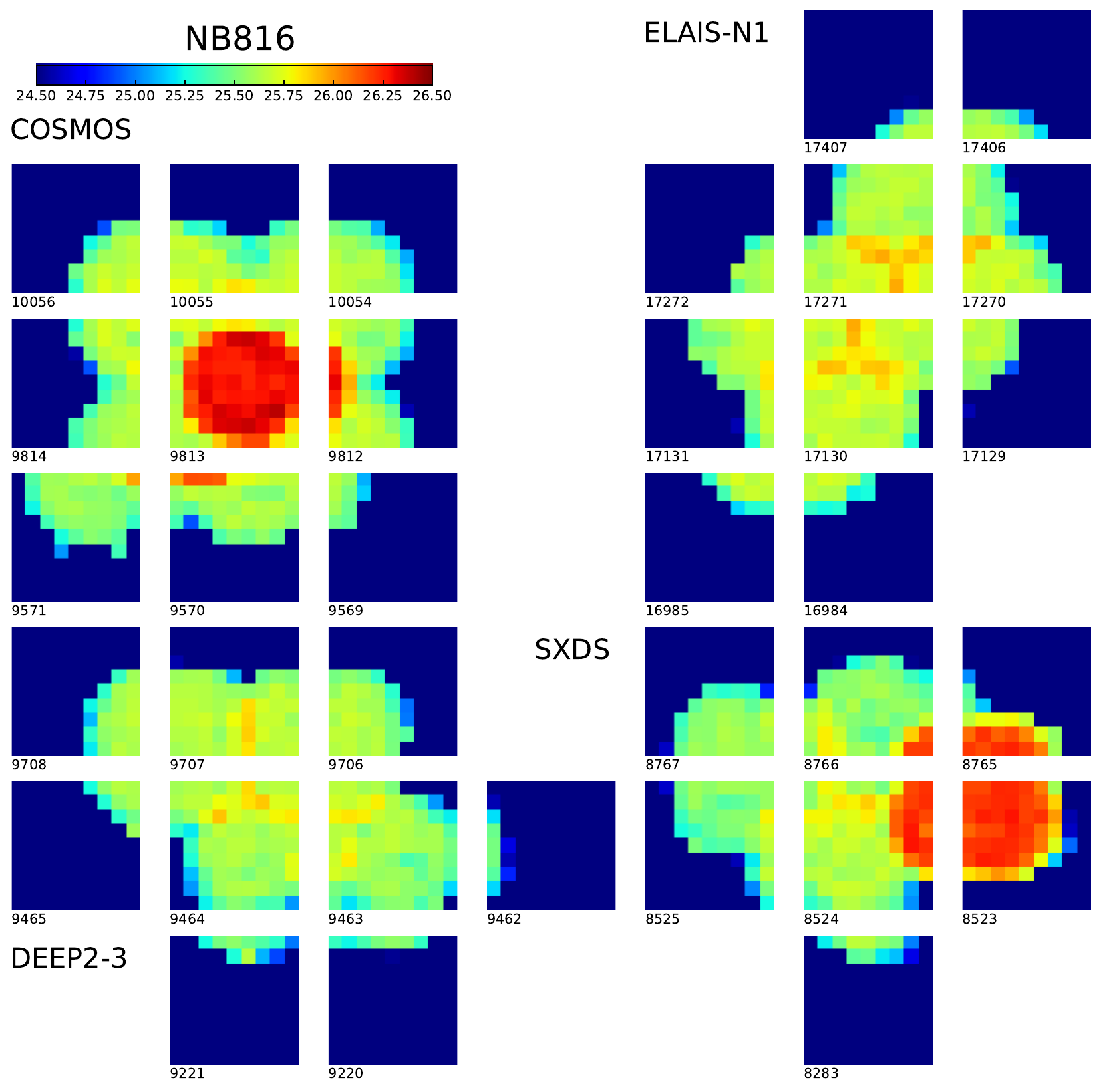}
    \caption{Limiting magnitude maps ($5\sigma$, 2{\mbox{$.\!\!\arcsec$}}0 aperture) for the NB816 images.}
    \label{fig:limmag0816}
\end{figure*}

\begin{figure*}[]
    \plotone{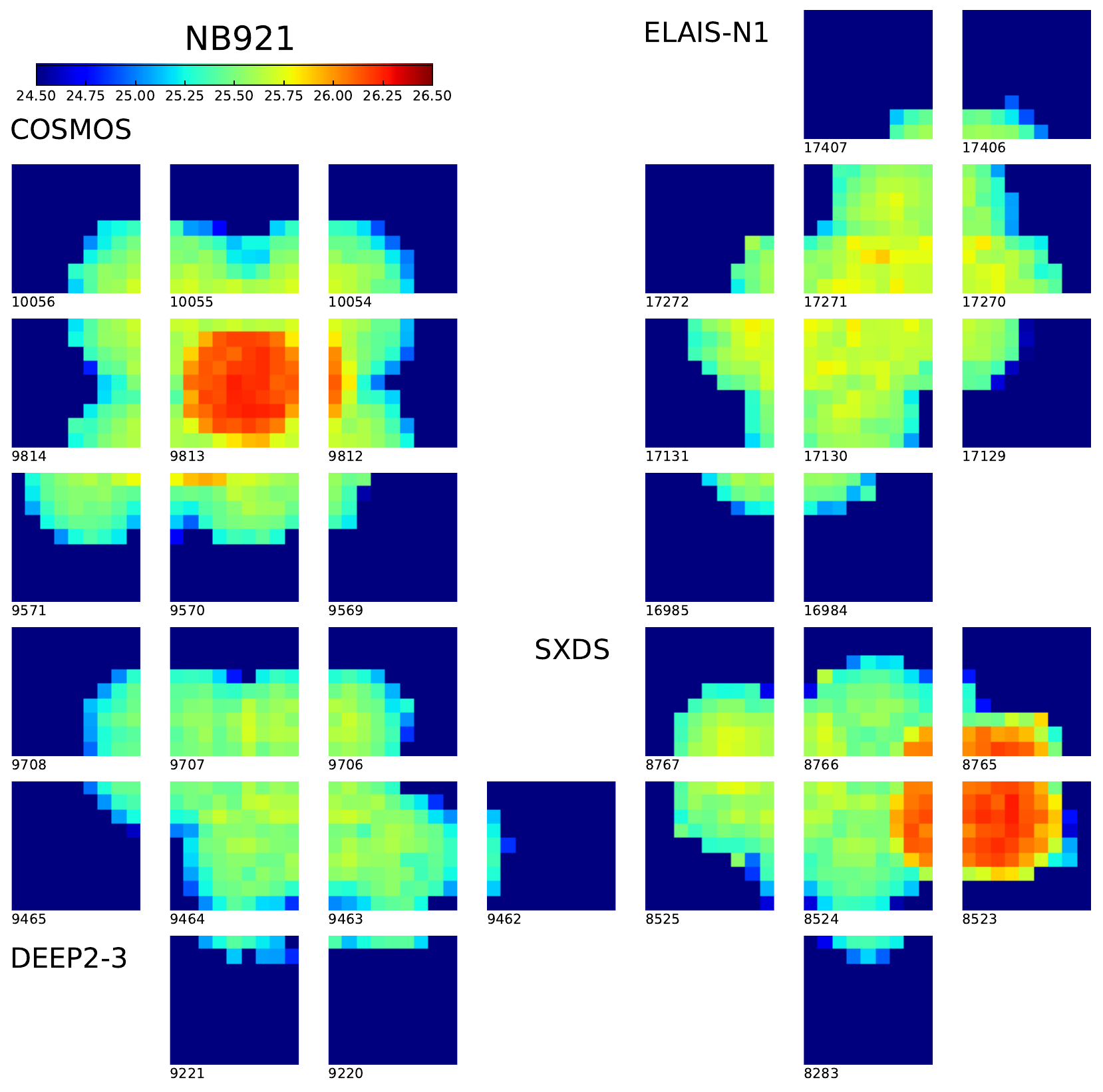}
    \caption{Limiting magnitude maps ($5\sigma$, 2{\mbox{$.\!\!\arcsec$}}0 aperture) for the NB921 images.}
    \label{fig:limmag0921}
\end{figure*}

\begin{figure*}[]
    \plotone{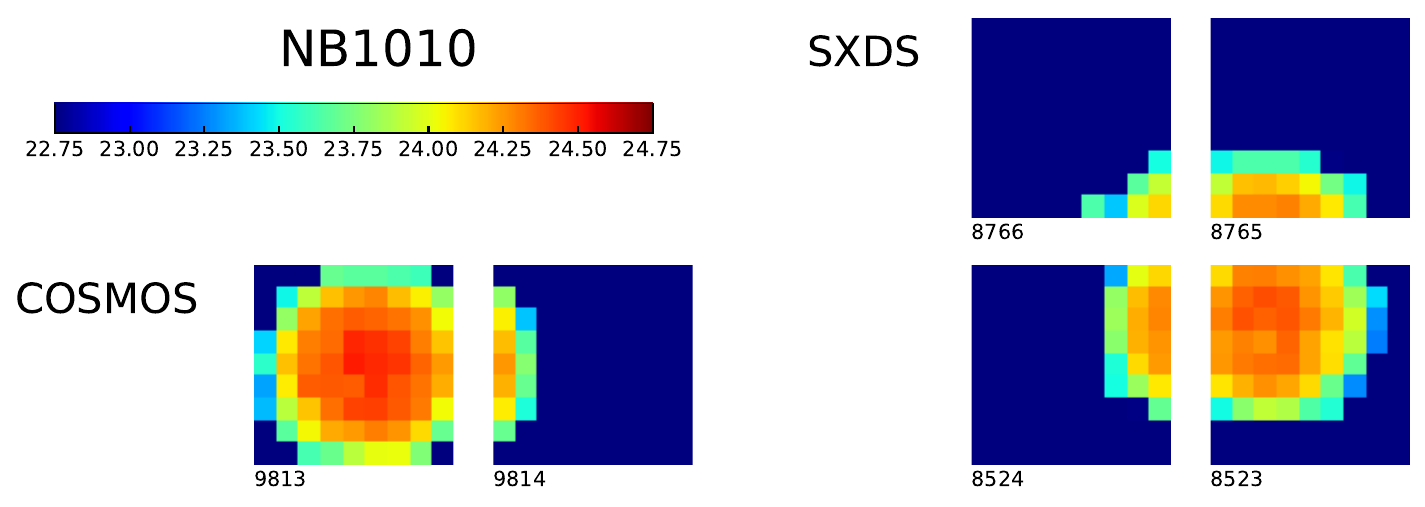}
    \caption{Limiting magnitude maps ($5\sigma$, 2{\mbox{$.\!\!\arcsec$}}0 aperture) for the NB1010 images.}
    \label{fig:limmag1010}
\end{figure*}

\begin{figure*}[]
    \plotone{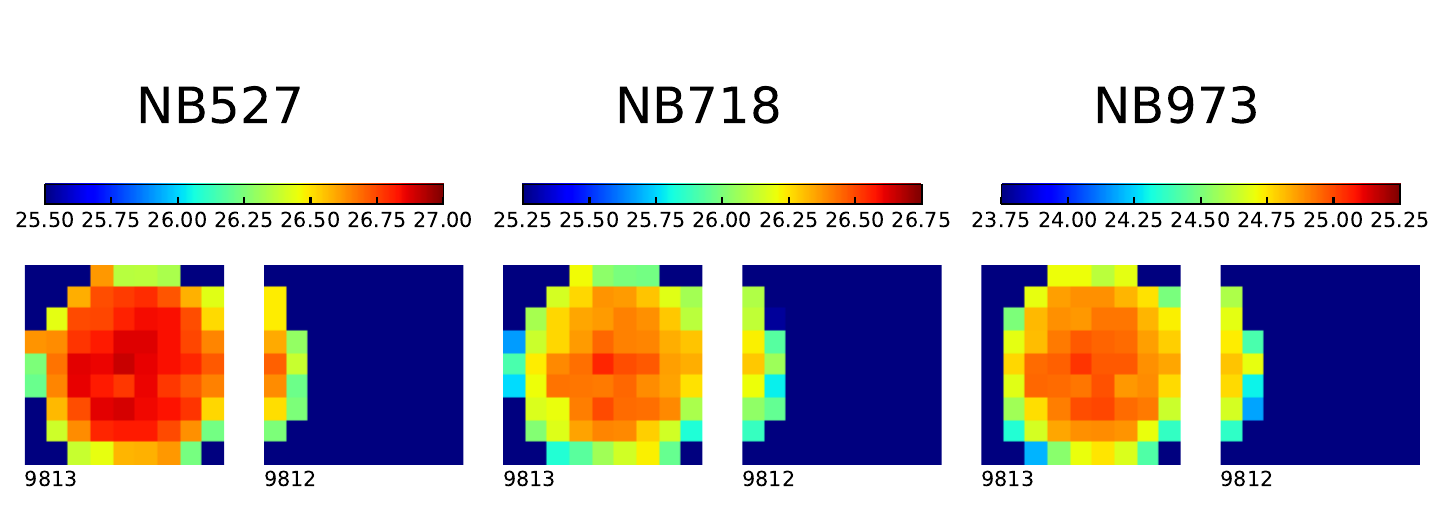}
    \caption{Limiting magnitude maps ($5\sigma$, 2{\mbox{$.\!\!\arcsec$}}0 aperture) for the CHORUS NBs (left: NB527; middle: NB718; right: NB973).}
    \label{fig:limmagchorus}
\end{figure*}

\begin{figure*}
    \plotone{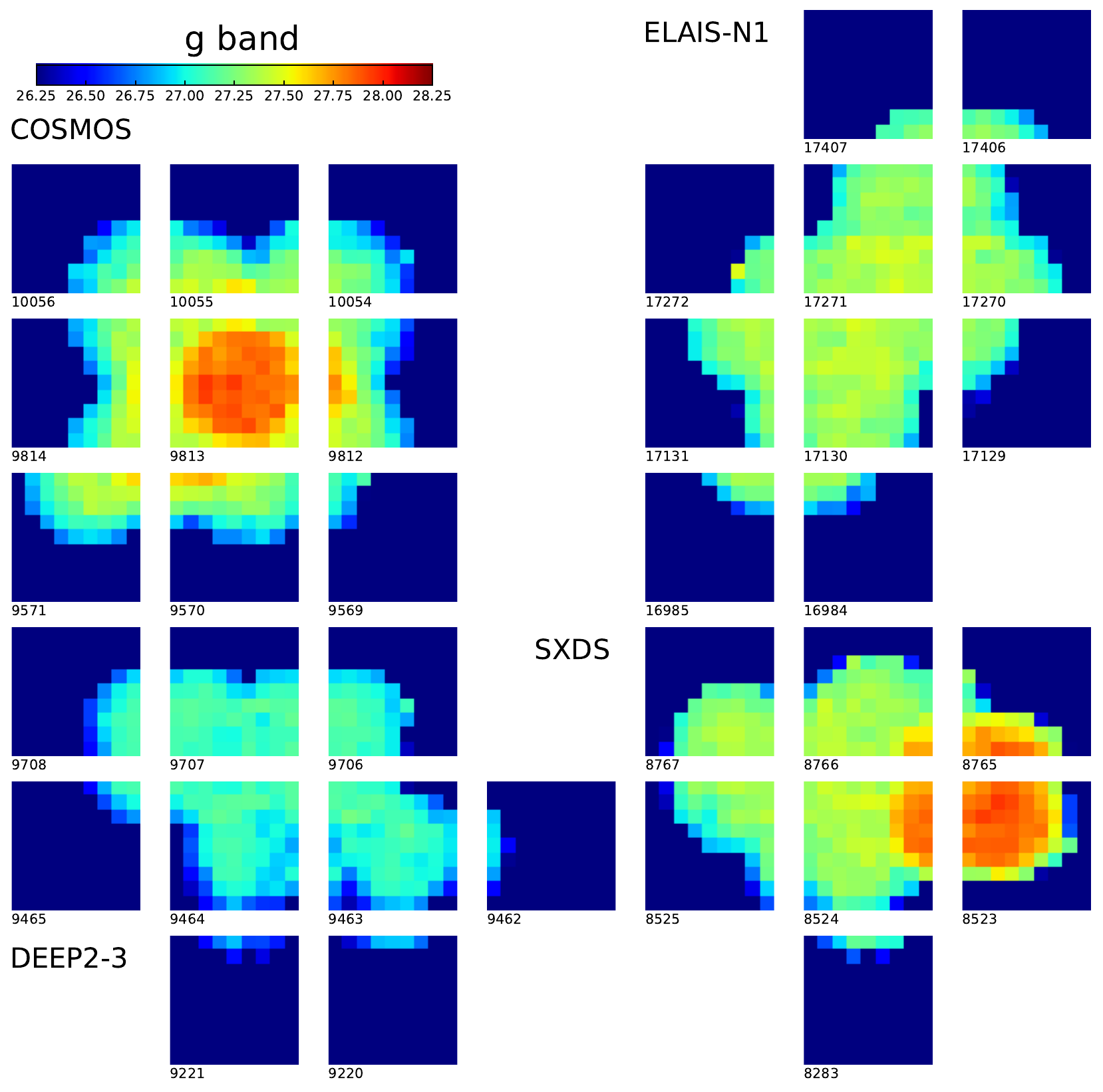}
    \caption{Limiting magnitude maps ($5\sigma$, 2{\mbox{$.\!\!\arcsec$}}0 aperture) for the g band.}
    \label{fig:limmagg}
\end{figure*}

\begin{figure*}
    \plotone{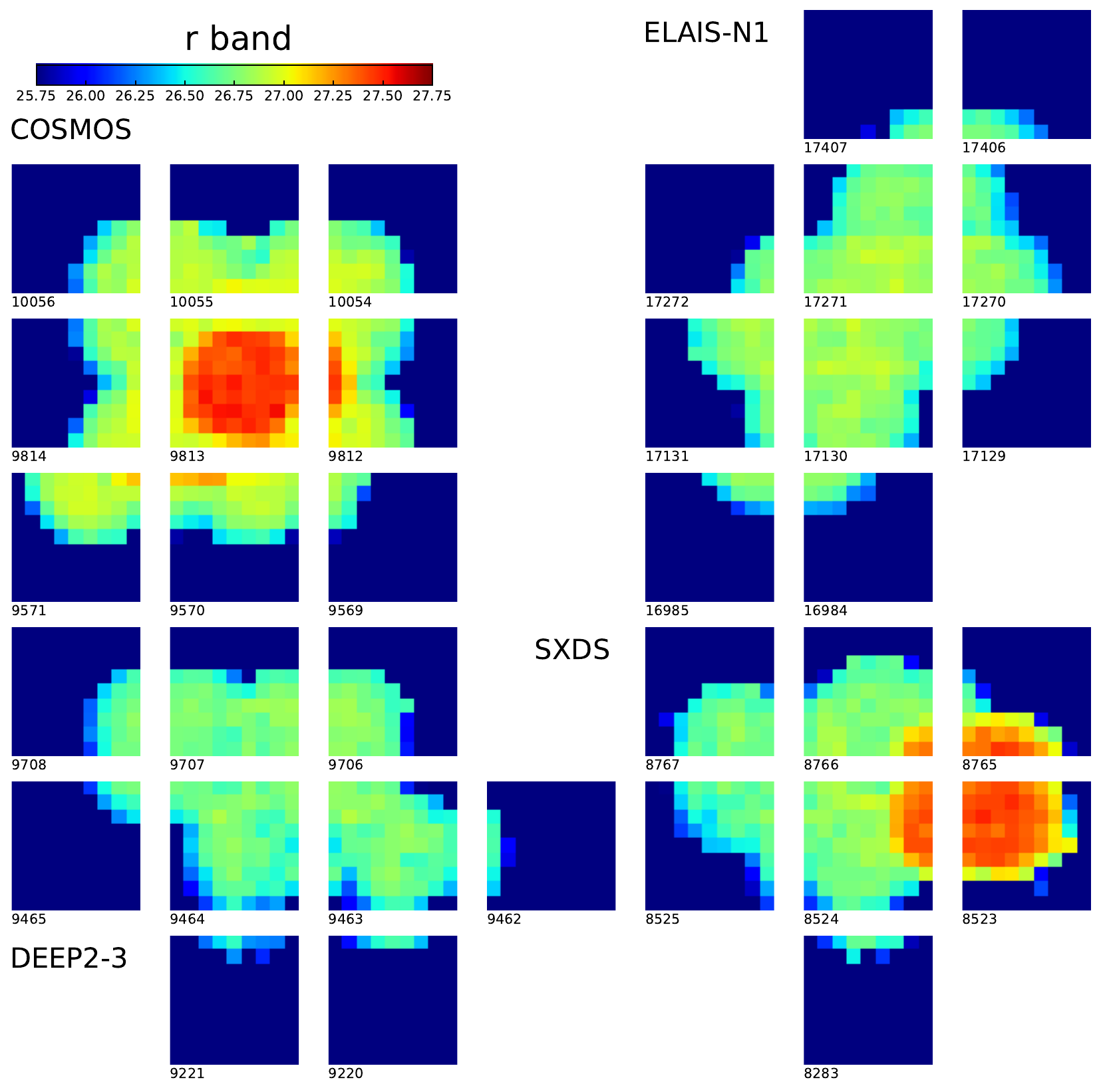}
    \caption{Limiting magnitude maps ($5\sigma$, 2{\mbox{$.\!\!\arcsec$}}0 aperture) for the r band.}
    \label{fig:limmagr}
\end{figure*}

\begin{figure*}
    \plotone{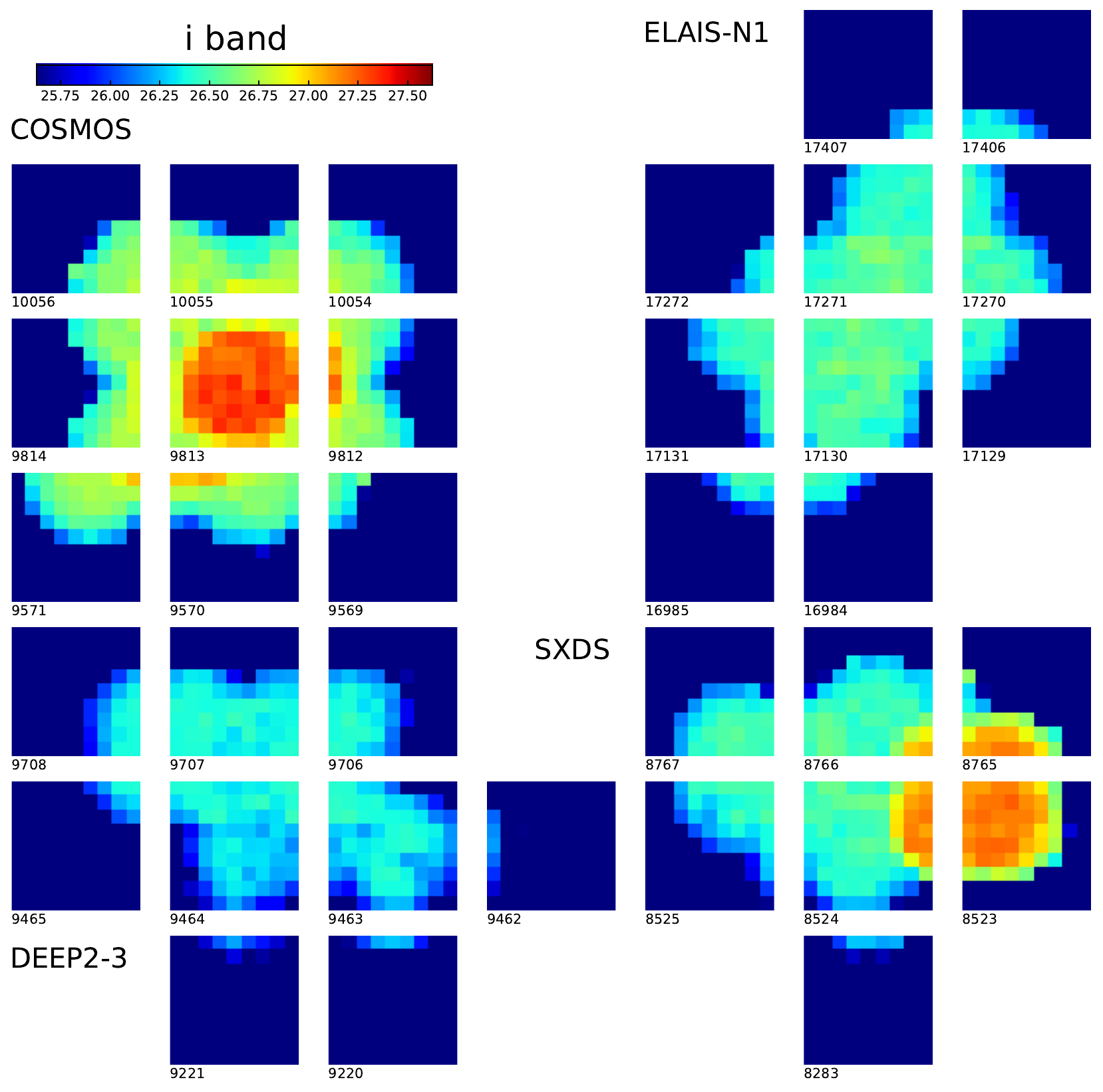}
    \caption{Limiting magnitude maps ($5\sigma$, 2{\mbox{$.\!\!\arcsec$}}0 aperture) for the i band.}
    \label{fig:limmagi}
\end{figure*}

\begin{figure*}
    \plotone{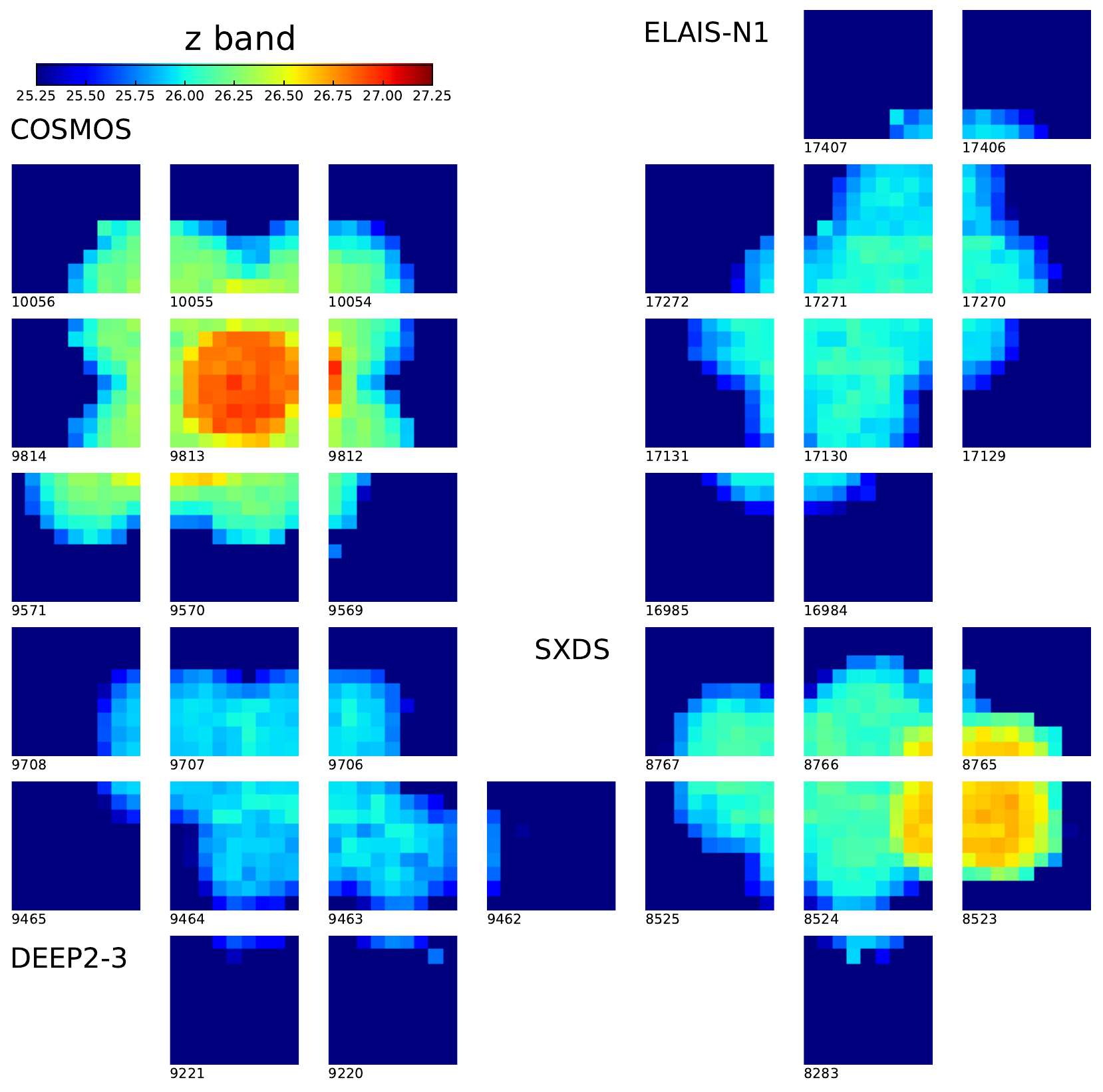}
    \caption{Limiting magnitude maps ($5\sigma$, 2{\mbox{$.\!\!\arcsec$}}0 aperture) for the z band.}
    \label{fig:limmagz}
\end{figure*}

\begin{figure*}
    \plotone{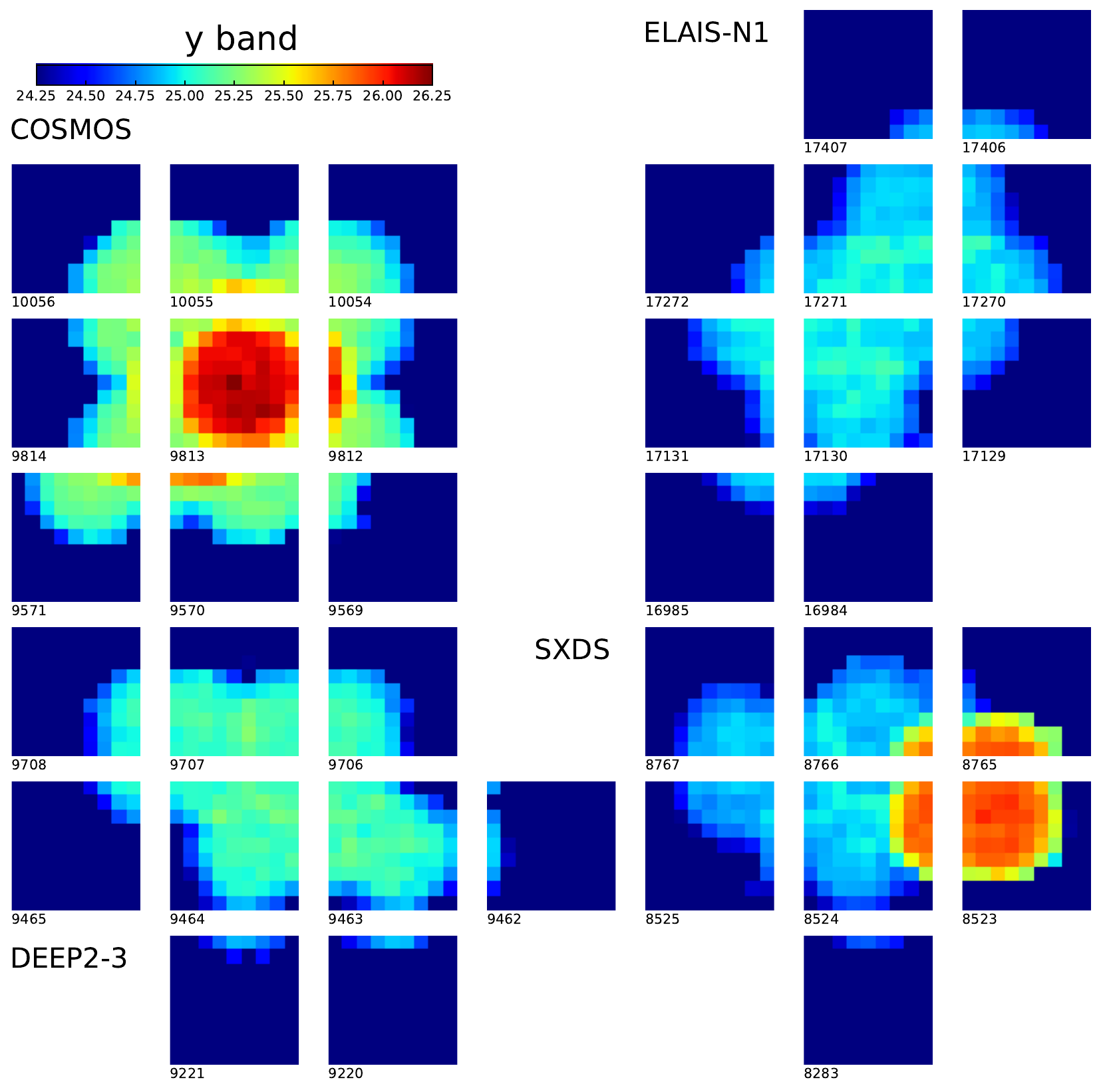}
    \caption{Limiting magnitude maps ($5\sigma$, 2{\mbox{$.\!\!\arcsec$}}0 aperture) for the y band.}
    \label{fig:limmagy}
\end{figure*}

Figures \ref{fig:limmag0387} -- \ref{fig:limmagy} show 5$\sigma$ limiting magnitude maps for the HSC filters we used in this work. 
The maps over the whole survey fields are shown for each tract, whose number is indicated at the bottom left. 
In each panel, $9\times9$ squares represent patches belonging to the tract. 
Aside from the D/UD distinction in COSMOS and SXDS, there are some variations that should be accounted for when LAEs are selected.



\section{List of Spectroscopically Confirmed Objects in Our Catalogs}\label{sec:ap:specmatch}

In Table \ref{tab:spec}, we list information (name, sky coordinates, spectroscopic redshift from each reference in the last column, photometric information (grizy and NB, 2{\mbox{$.\!\!\arcsec$}}0-aperture magnitudes) in our images, LAE sample in which the source is included, notable features (if any), and reference) on our LAEs confirmed by past spectroscopic observations.
We also list information on sources identified as lower-redshift objects in spectroscopic catalogs in Table \ref{tab:lowzspec}.

\begin{longrotatetable}
\begin{deluxetable*}{lcccccccccccc}
\tablecaption{List of Objects with Spectroscopic Confirmation in Our LAE Catalogs\label{tab:spec}}
\tablehead{
\colhead{ID} & \colhead{R.A.} & \colhead{Decl.} & \colhead{$z_\mathrm{spec}$} & \colhead{$\mathrm{g_{ap}}$} & \colhead{$\mathrm{r_{ap}}$} & \colhead{$\mathrm{i_{ap}}$} & \colhead{$\mathrm{z_{ap}}$} & \colhead{$\mathrm{y_{ap}}$} & \colhead{$\mathrm{NB_{ap}}$} & \colhead{Sample} & \colhead{Feature} & \colhead{Reference}  \\ 
\colhead{(1)} & \colhead{(2)} & \colhead{(3)} & \colhead{(4)} & 
\colhead{(5)} & \colhead{(6)} & \colhead{(7)} &
\colhead{(8)} & \colhead{(9)} & \colhead{(10)} & \colhead{(11)} & \colhead{(12)} & \colhead{(13)}
} 
\startdata
HSC J022013--050729 & 02:20:13.03 & --05:07:29.7 & 2.169 & 22.7 & 22.7 & 22.6 & 22.3 & 22.4 & 21.7 & NB387 & AGN & Co11 \\
HSC J022028--045802 & 02:20:28.96 & --04:58:02.8 & 2.157 & 20.4 & 20.5 & 20.3 & 20.0 & 20.1 & 19.1 & NB387 & AGN & Co11, Ly20 \\
HSC J022044--050906 & 02:20:44.46 & --05:09:06.1 & 2.105 & 20.8 & 20.8 & 20.8 & 20.5 & 20.6 & 20.0 & NB387 & AGN & Li15, Sc18, Ly20 \\
HSC J022120--032353 & 02:21:20.93 & --03:23:53.5 & 2.183 & 22.6 & 22.5 & 22.4 & 22.2 & 22.3 & 21.2 & NB387 & AGN & Ly20 \\
HSC J022121--034338 & 02:21:21.16 & --03:43:38.1 & 2.176 & 23.1 & 23.2 & 23.0 & 22.7 & 22.7 & 21.9 & NB387 & AGN & Ly20 \\
HSC J022202--050352 & 02:22:02.59 & --05:03:52.0 & 2.161 & 21.6 & 21.6 & 21.6 & 21.3 & 21.5 & 20.4 & NB387 & AGN & Ly20 \\
HSC J022219--052231 & 02:22:19.96 & --05:22:31.5 & 2.204 & 20.4 & 20.3 & 20.3 & 20.0 & 20.1 & 19.3 & NB387 & AGN & Li15, Ly20 \\
HSC J022230--033545 & 02:22:30.42 & --03:35:45.4 & 2.172 & 22.1 & 22.4 & 22.1 & 21.7 & 21.9 & 20.0 & NB387 & AGN & Ly20 \\
HSC J022312--050625 & 02:23:12.46 & --05:06:25.1 & 2.190 & 20.1 & 20.1 & 20.1 & 19.9 & 19.9 & 19.1 & NB387 & AGN & Ly20 \\
HSC J022349--033930 & 02:23:49.43 & --03:39:30.6 & 2.141 & 20.9 & 20.8 & 20.8 & 20.5 & 20.7 & 20.0 & NB387 & AGN & Ly20 \\
HSC J022351--044730 & 02:23:51.07 & --04:47:30.0 & 2.167 & 20.7 & 20.6 & 20.6 & 20.3 & 20.4 & 19.6 & NB387 & AGN & Ly20 \\
HSC J022435--053615 & 02:24:35.54 & --05:36:15.6 & 2.217 & 22.0 & 22.0 & 21.9 & 21.6 & 21.7 & 21.1 & NB387 & AGN & Li15, Ly20 \\
HSC J022548--041028 & 02:25:48.98 & --04:10:28.4 & 2.124 & 23.8 & 23.7 & 23.5 & 23.1 & 23.1 & 22.5 & NB387 &  & Li15 \\
HSC J022722--045221 & 02:27:22.20 & --04:52:21.9 & 2.127 & 21.6 & 21.7 & 21.6 & 21.3 & 21.5 & 20.7 & NB387 & AGN & Ly20 \\
HSC J022751--050101 & 02:27:51.25 & --05:01:01.3 & 2.175 & 21.6 & 21.9 & 21.8 & 21.3 & 21.5 & 20.6 & NB387 & AGN & Li15, Ly20 \\
HSC J095508+011205 & 09:55:08.45 & +01:12:05.7 & 2.166 & 21.8 & 21.9 & 21.6 & 21.3 & 21.5 & 19.9 & NB387 & AGN & Ly20 \\
HSC J095539+011316 & 09:55:39.50 & +01:13:16.4 & 2.142 & 21.6 & 21.6 & 21.6 & 21.1 & 21.4 & 20.5 & NB387 & AGN & Ly20 \\
HSC J095714+013145 & 09:57:14.02 & +01:31:45.5 & 2.142 & 21.7 & 21.6 & 21.4 & 21.0 & 21.1 & 20.1 & NB387 & AGN & Ly20 \\
HSC J095822+010806 & 09:58:22.03 & +01:08:06.4 & 2.171 & 20.7 & 20.6 & 20.6 & 20.3 & 20.4 & 19.4 & NB387 & AGN & Ly20 \\
HSC J095944+020742 & 09:59:44.25 & +02:07:42.7 & 2.191 & 24.8 & 24.8 & 24.9 & 24.8 & 24.9 & 23.6 & NB387 &  & Sh14 \\
HSC J095945+022808 & 09:59:45.98 & +02:28:08.0 & 2.174 & 24.9 & 24.8 & 24.8 & 24.7 & 24.9 & 23.7 & NB387 &  & Sh14 \\
HSC J095954+022630 & 09:59:54.39 & +02:26:30.0 & 2.194 & 24.9 & 24.6 & 24.3 & 23.9 & 23.7 & 23.3 & NB387 &  & Sh14 \\
HSC J095950+020347 & 09:59:50.74 & +02:03:47.7 & 2.184 & 24.5 & 24.5 & 24.5 & 24.4 & 24.4 & 23.2 & NB387 &  & Has18 \\
HSC J095959+020531 & 09:59:59.06 & +02:05:31.6 & 2.182 & 25.7 & 25.6 & 25.6 & 25.6 & 25.8 & 24.6 & NB387 &  & Sh14 \\
HSC J095959+020838 & 09:59:59.38 & +02:08:38.4 & 2.163 & 24.5 & 24.5 & 24.5 & 24.5 & 24.5 & 22.9 & NB387 &  & Sh14 \\
HSC J100012+021750 & 10:00:12.56 & +02:17:50.4 & 2.161 & 24.4 & 24.3 & 24.3 & 24.3 & 24.4 & 23.3 & NB387 &  & Kr15 \\
HSC J100015+020807 & 10:00:15.26 & +02:08:07.4 & 2.161 & 24.7 & 24.6 & 24.6 & 24.6 & 24.6 & 23.8 & NB387 &  & Sh14 \\
HSC J100017+030525 & 10:00:17.73 & +03:05:25.0 & 2.174 & 20.7 & 20.9 & 20.4 & 20.1 & 20.4 & 18.8 & NB387 & AGN & Ly20 \\
HSC J100025+022357 & 10:00:25.98 & +02:23:57.4 & 2.169 & 23.9 & 23.8 & 23.7 & 23.6 & 23.4 & 23.1 & NB387 &  & Mo16 \\
HSC J100032+021535 & 10:00:32.66 & +02:15:35.6 & 2.157 & 26.5 & 26.2 & 26.2 & 26.2 & 26.5 & 25.5 & NB387 &  & Kr15, Mo16 \\
HSC J100047+020757 & 10:00:47.75 & +02:07:57.0 & 2.153 & 19.9 & 19.8 & 19.7 & 19.4 & 19.5 & 18.8 & NB387 & AGN & Li09, Ly20 \\
HSC J100102+011948 & 10:01:02.16 & +01:19:48.6 & 2.146 & 19.7 & 19.7 & 19.6 & 19.4 & 19.6 & 18.8 & NB387 & AGN & Ly20 \\
HSC J100146+033040 & 10:01:46.03 & +03:30:40.1 & 2.212 & 22.4 & 22.5 & 22.2 & 21.7 & 22.2 & 21.2 & NB387 & AGN & Ly20 \\
HSC J160328+545715 & 16:03:28.75 & +54:57:15.0 & 2.142 & 21.2 & 21.3 & 21.1 & 20.9 & 21.0 & 20.2 & NB387 & AGN & Ly20 \\
HSC J160554+544859 & 16:05:54.66 & +54:48:59.5 & 2.159 & 21.4 & 21.6 & 21.5 & 21.2 & 21.4 & 20.0 & NB387 & AGN & Ly20 \\
HSC J160611+542447 & 16:06:11.62 & +54:24:47.5 & 2.198 & 20.4 & 20.5 & 20.4 & 20.1 & 20.3 & 19.5 & NB387 & AGN & Ly20 \\
HSC J161105+532237 & 16:11:05.06 & +53:22:37.9 & 2.182 & 21.6 & 21.8 & 21.6 & 21.2 & 21.5 & 20.1 & NB387 & AGN & Ly20 \\
HSC J161129+545055 & 16:11:29.04 & +54:50:55.5 & 2.108 & 20.4 & 20.4 & 20.3 & 20.1 & 20.2 & 19.6 & NB387 & AGN & Ly20 \\
HSC J161327+541256 & 16:13:27.44 & +54:12:56.2 & 2.155 & 21.3 & 21.3 & 21.1 & 20.8 & 20.9 & 20.0 & NB387 & AGN & Ly20 \\
HSC J161432+543903 & 16:14:32.77 & +54:39:03.7 & 2.193 & 21.6 & 21.7 & 21.5 & 21.2 & 21.4 & 20.1 & NB387 & AGN & Ly20 \\
HSC J161552+553101 & 16:15:52.16 & +55:31:01.9 & 2.160 & 21.2 & 21.5 & 21.3 & 21.1 & 21.2 & 20.0 & NB387 & AGN & Ly20 \\
HSC J232147--010038 & 23:21:47.17 & --01:00:38.9 & 2.177 & 21.3 & 21.2 & 21.0 & 20.7 & 20.7 & 20.1 & NB387 & AGN & Ly20 \\
HSC J232459+001451 & 23:24:59.70 & +00:14:51.2 & 2.172 & 21.3 & 21.2 & 21.1 & 20.8 & 20.9 & 19.6 & NB387 & AGN & Ly20 \\
HSC J232506--012203 & 23:25:06.73 & --01:22:03.4 & 2.181 & 20.3 & 20.3 & 20.3 & 20.1 & 20.2 & 19.3 & NB387 & AGN & Ly20 \\
HSC J232608--011540 & 23:26:08.69 & --01:15:40.3 & 2.194 & 22.7 & 22.6 & 22.3 & 21.8 & 21.9 & 21.1 & NB387 & AGN & Ly20 \\
HSC J232614+003249 & 23:26:14.28 & +00:32:49.2 & 2.180 & 20.8 & 20.5 & 20.4 & 20.2 & 20.2 & 19.6 & NB387 & AGN & Ly20 \\
HSC J232619+000152 & 23:26:19.36 & +00:01:52.7 & 2.173 & 22.1 & 22.1 & 21.9 & 21.5 & 21.7 & 21.0 & NB387 & AGN & Ly20 \\
HSC J232649--000010 & 23:26:49.36 & --00:00:10.1 & 2.168 & 22.7 & 22.6 & 22.3 & 21.7 & 21.9 & 21.1 & NB387 & AGN & Ly20 \\
HSC J232855--004212 & 23:28:55.78 & --00:42:12.2 & 2.198 & 20.5 & 20.4 & 20.1 & 19.8 & 19.9 & 19.0 & NB387 & AGN & Ly20 \\
HSC J233159+000856 & 23:31:59.69 & +00:08:56.5 & 2.184 & 20.9 & 20.7 & 20.7 & 20.3 & 20.4 & 19.5 & NB387 & AGN & Co11, Ly20 \\
HSC J095759+015804 & 09:57:59.01 & +01:58:04.7 & 3.261 & 23.7 & 23.3 & 23.5 & 23.5 & 23.4 & 22.6 & NB527 & AGN & Has18 \\
HSC J095940+025255 & 09:59:40.18 & +02:52:55.4 & 3.317 & 23.9 & 23.6 & 23.5 & 23.4 & 23.4 & 22.5 & NB527 & AGN & Has18 \\
HSC J100013+013704 & 10:00:13.54 & +01:37:04.6 & 3.489 & 27.0 & 26.8 & 26.2 & 26.0 & 26.1 & 24.5 & NB527 &  & Has18 \\
HSC J100017+021657 & 10:00:17.38 & +02:16:57.5 & 3.295 & 25.6 & 25.1 & 24.8 & 24.7 & 24.5 & 23.9 & NB527 &  & Ta17 \\
HSC J100033+021825 & 10:00:33.33 & +02:18:25.3 & 3.288 & 25.2 & 24.8 & 24.6 & 24.6 & 24.6 & 24.1 & NB527 &  & Ta17 \\
HSC J100057+023932 & 10:00:57.80 & +02:39:32.6 & 3.360 & 23.5 & 22.9 & 22.8 & 22.8 & 22.8 & 22.3 & NB527 & AGN & Co11, Mas12 \\
HSC J100113+014541 & 10:01:13.46 & +01:45:41.8 & 3.310 & 25.9 & 25.4 & 25.2 & 25.0 & 25.0 & 24.0 & NB527 &  & Has18 \\
HSC J100220+020453 & 10:02:20.39 & +02:04:53.0 & 3.326 & 24.3 & 24.1 & 23.8 & 23.6 & 23.3 & 22.3 & NB527 & AGN & Co11, Has18 \\
HSC J095820+021658 & 09:58:20.94 & +02:16:58.5 & 4.909 & 28.7 & 27.5 & 25.5 & 25.3 & 25.8 & 24.0 & NB718 & AGN & Ma12, Has18 \\
HSC J095830+020631 & 09:58:30.61 & +02:06:31.2 & 4.891 & 99.0 & 27.3 & 25.0 & 24.9 & 25.0 & 24.1 & NB718 &  & Ma12, Has18 \\
HSC J095929+022950 & 09:59:29.35 & +02:29:50.2 & 4.841 & 99.0 & 27.9 & 25.7 & 26.0 & 26.0 & 24.9 & NB718 &  & Ma12, Has18 \\
HSC J095950+022451 & 09:59:50.03 & +02:24:51.5 & 4.843 & 99.0 & 28.5 & 26.6 & 27.4 & 28.6 & 25.9 & NB718 &  & Ma12, Has18 \\
HSC J095956+015451 & 09:59:56.14 & +01:54:51.5 & 4.844 & 99.0 & 29.4 & 26.5 & 27.4 & 28.0 & 25.6 & NB718 &  & Ma12, Has18 \\
HSC J100004+020845 & 10:00:04.17 & +02:08:45.6 & 4.840 & 99.0 & 26.6 & 25.0 & 25.2 & 25.2 & 24.2 & NB718 &  & Ma12, Has18 \\
HSC J100008+023456 & 10:00:08.76 & +02:34:56.5 & 4.868 & 99.0 & 28.2 & 26.7 & 26.8 & 26.4 & 25.7 & NB718 &  & Has18 \\
HSC J100030+013621 & 10:00:30.43 & +01:36:21.7 & 4.844 & 29.2 & 27.5 & 25.7 & 26.0 & 25.9 & 25.0 & NB718 &  & Ma12, Has18 \\
HSC J100041+022637 & 10:00:41.08 & +02:26:37.4 & 4.869 & 99.0 & 27.5 & 25.6 & 25.8 & 25.9 & 24.5 & NB718 &  & Ma12, Has18 \\
HSC J100055+021309 & 10:00:55.43 & +02:13:09.1 & 4.873 & 99.0 & 28.6 & 26.1 & 26.3 & 27.3 & 24.1 & NB718 &  & Ma12, Has18 \\
HSC J100122+022249 & 10:01:22.45 & +02:22:49.8 & 4.865 & 29.4 & 28.5 & 26.4 & 26.5 & 26.3 & 24.9 & NB718 &  & Ma12, Has18 \\
HSC J100145+015712 & 10:01:45.12 & +01:57:12.2 & 4.910 & 99.0 & 26.6 & 25.1 & 25.0 & 24.9 & 24.4 & NB718 &  & Ma12, Has18 \\
HSC J100145+020244 & 10:01:45.96 & +02:02:44.3 & 4.888 & 30.0 & 25.8 & 24.1 & 24.1 & 24.1 & 23.0 & NB718 & LAB & Zh20 \\
HSC J021524--050250 & 02:15:24.81 & --05:02:50.9 & 5.702 & 99.0 & 28.7 & 27.1 & 27.7 & 28.3 & 25.4 & NB816 &  & Ni20 \\
HSC J021525--045918 & 02:15:25.25 & --04:59:18.2 & 5.674 & 99.0 & 99.0 & 26.6 & 25.8 & 25.7 & 24.8 & NB816 &  & Ji17, Ni20 \\
HSC J021526--045229 & 02:15:26.23 & --04:52:29.9 & 5.655 & 99.0 & 99.0 & 26.1 & 25.2 & 25.0 & 24.7 & NB816 &  & On21 \\
HSC J021533--050137 & 02:15:33.25 & --05:01:37.3 & 5.671 & 29.5 & 28.2 & 26.4 & 25.4 & 25.3 & 24.7 & NB816 &  & On21 \\
HSC J021551--045325 & 02:15:51.35 & --04:53:25.5 & 5.712 & 99.0 & 99.0 & 27.8 & 26.9 & 27.2 & 25.3 & NB816 &  & Ni20, On21 \\
HSC J021553--051000 & 02:15:53.98 & --05:10:00.4 & 5.732 & 99.0 & 29.0 & 99.0 & 28.7 & 29.0 & 25.7 & NB816 &  & This Study \\
HSC J021555--045318 & 02:15:55.66 & --04:53:18.9 & 5.740 & 99.0 & 28.8 & 27.2 & 27.6 & 27.3 & 25.3 & NB816 &  & Ni20, On21 \\
HSC J021558--045301 & 02:15:58.50 & --04:53:01.9 & 5.718 & 99.0 & 99.0 & 27.2 & 27.6 & 27.8 & 25.0 & NB816 &  & On21 \\
HSC J021601--045404 & 02:16:01.36 & --04:54:04.2 & 5.722 & 99.0 & 99.0 & 99.0 & 29.4 & 99.0 & 25.9 & NB816 &  & Ni20 \\
HSC J021609--045651 & 02:16:09.47 & --04:56:51.3 & 5.691 & 99.0 & 99.0 & 28.4 & 99.0 & 99.0 & 25.8 & NB816 &  & Ni20, This Study \\
HSC J021612--051308 & 02:16:12.32 & --05:13:08.5 & 5.724 & 99.0 & 99.0 & 28.1 & 99.0 & 99.0 & 25.4 & NB816 &  & Ni20 \\
HSC J021617--045419 & 02:16:17.16 & --04:54:19.4 & 5.710 & 99.0 & 99.0 & 28.5 & 29.1 & 27.6 & 25.6 & NB816 &  & Ni20, On21 \\
HSC J021624--045516 & 02:16:24.71 & --04:55:16.7 & 5.707 & 99.0 & 99.0 & 26.4 & 25.9 & 25.8 & 23.9 & NB816 &  & Ji17, Ni20 \\
HSC J021625--045237 & 02:16:25.65 & --04:52:37.3 & 5.730 & 99.0 & 99.0 & 28.2 & 28.9 & 27.1 & 25.1 & NB816 &  & On21 \\
HSC J021628--050103 & 02:16:28.06 & --05:01:03.8 & 5.695 & 99.0 & 29.2 & 27.5 & 28.2 & 99.0 & 25.1 & NB816 &  & Ni20, On21 \\
HSC J021629--050259 & 02:16:29.60 & --05:02:59.2 & 5.678 & 29.1 & 99.0 & 27.9 & 27.9 & 28.5 & 25.9 & NB816 &  & Ni20 \\
HSC J021630--045734 & 02:16:30.40 & --04:57:34.4 & 5.700 & 29.8 & 29.0 & 28.0 & 99.0 & 28.3 & 25.9 & NB816 &  & This Study \\
HSC J021636--044723 & 02:16:36.45 & --04:47:23.7 & 5.721 & 99.0 & 29.9 & 27.2 & 26.5 & 26.7 & 25.2 & NB816 &  & Ni20, On21 \\
HSC J021639--051346 & 02:16:39.90 & --05:13:46.9 & 5.705 & 99.0 & 99.0 & 27.6 & 27.4 & 27.7 & 25.3 & NB816 &  & Ni20 \\
HSC J021654--052155 & 02:16:54.61 & --05:21:55.7 & 5.715 & 29.0 & 28.7 & 26.8 & 26.9 & 26.7 & 24.7 & NB816 &  & Ji18, Ha19, Ni20 \\
HSC J021657--052117 & 02:16:57.89 & --05:21:17.3 & 5.669 & 99.0 & 99.0 & 26.7 & 26.4 & 27.2 & 25.3 & NB816 &  & Ji17, Ji18, Ha19, Ni20 \\
HSC J021659--052305 & 02:16:59.95 & --05:23:05.4 & 5.700 & 29.0 & 99.0 & 28.1 & 27.9 & 27.1 & 25.7 & NB816 &  & Ha19 \\
HSC J021700--051810 & 02:17:00.80 & --05:18:10.5 & 5.713 & 99.0 & 99.0 & 28.0 & 27.9 & 99.0 & 25.6 & NB816 &  & Ji18, Ni20 \\
HSC J021700--053130 & 02:17:00.62 & --05:31:30.5 & 5.754 & 99.0 & 29.6 & 27.0 & 26.9 & 26.5 & 25.2 & NB816 &  & Ji18, Ni20 \\
HSC J021701--051841 & 02:17:01.44 & --05:18:41.6 & 5.679 & 99.0 & 99.0 & 28.0 & 27.1 & 27.6 & 25.7 & NB816 &  & Ha19 \\
HSC J021704--052714 & 02:17:04.30 & --05:27:14.4 & 5.687 & 99.0 & 28.4 & 26.5 & 26.6 & 26.1 & 24.4 & NB816 &  & Ji17, Ji18, Ha19, Ni20 \\
HSC J021705--052427 & 02:17:05.19 & --05:24:27.6 & 5.705 & 99.0 & 29.7 & 27.9 & 99.0 & 99.0 & 26.0 & NB816 &  & Ni20 \\
HSC J021707--052723 & 02:17:07.98 & --05:27:23.4 & 5.720 & 99.0 & 99.0 & 27.4 & 28.7 & 99.0 & 26.1 & NB816 &  & Ha19 \\
HSC J021707--053426 & 02:17:07.86 & --05:34:26.7 & 5.680 & 99.0 & 99.0 & 26.5 & 26.3 & 26.4 & 24.4 & NB816 &  & Ji17, Ji18, Ha19, Ni20 \\
HSC J021708--052421 & 02:17:08.87 & --05:24:21.7 & 5.675 & 29.3 & 28.5 & 27.2 & 26.9 & 27.4 & 25.7 & NB816 &  & Ji18, Ni20 \\
HSC J021709--050329 & 02:17:09.78 & --05:03:29.3 & 5.709 & 99.0 & 29.3 & 27.1 & 26.7 & 28.9 & 25.0 & NB816 &  & On21 \\
HSC J021709--052646 & 02:17:09.96 & --05:26:46.7 & 5.689 & 99.0 & 99.0 & 27.8 & 27.0 & 27.5 & 25.4 & NB816 &  & Ha19, Ni20 \\
HSC J021709--052731 & 02:17:09.52 & --05:27:31.5 & 5.676 & 99.0 & 29.6 & 27.8 & 99.0 & 28.3 & 25.7 & NB816 &  & Ji18, Ha19, Ni20 \\
HSC J021710--044351 & 02:17:10.47 & --04:43:51.6 & 5.691 & 99.0 & 28.0 & 27.4 & 28.6 & 27.4 & 25.9 & NB816 &  & Ni20, This Study \\
HSC J021713--053558 & 02:17:13.80 & --05:35:58.0 & 5.686 & 28.7 & 99.0 & 27.4 & 26.0 & 26.0 & 25.2 & NB816 &  & Ha19 \\
HSC J021714--053502 & 02:17:14.93 & --05:35:02.9 & 5.685 & 99.0 & 29.9 & 27.7 & 26.3 & 29.7 & 25.5 & NB816 &  & Ha19 \\
HSC J021719--043150 & 02:17:19.14 & --04:31:50.7 & 5.735 & 99.0 & 99.0 & 27.5 & 26.8 & 27.2 & 25.0 & NB816 &  & Ni20, On21 \\
HSC J021722--052805 & 02:17:22.28 & --05:28:05.4 & 5.682 & 99.0 & 99.0 & 27.6 & 27.6 & 27.0 & 25.7 & NB816 &  & Ji18, Ha19, Ni20 \\
HSC J021723--052559 & 02:17:23.53 & --05:25:59.9 & 5.707 & 99.0 & 99.0 & 28.9 & 29.1 & 27.6 & 25.8 & NB816 &  & This Study \\
HSC J021724--053309 & 02:17:24.04 & --05:33:09.8 & 5.708 & 29.0 & 28.2 & 25.8 & 25.5 & 25.8 & 23.6 & NB816 &  & Ji17, Ji18, Sh18, Ha19, Ni20 \\
HSC J021725--050737 & 02:17:25.90 & --05:07:37.7 & 5.701 & 99.0 & 99.0 & 27.7 & 27.0 & 26.7 & 24.8 & NB816 &  & On21 \\
HSC J021726--045126 & 02:17:26.74 & --04:51:26.8 & 5.709 & 99.0 & 99.0 & 27.6 & 99.0 & 28.3 & 25.0 & NB816 &  & On21 \\
HSC J021729--053028 & 02:17:29.19 & --05:30:28.6 & 5.746 & 99.0 & 99.0 & 27.8 & 28.8 & 99.0 & 25.1 & NB816 &  & Ha19 \\
HSC J021733--050901 & 02:17:33.18 & --05:09:01.3 & 5.679 & 99.0 & 99.0 & 27.7 & 29.7 & 28.6 & 25.7 & NB816 &  & Ni20 \\
HSC J021734--053452 & 02:17:34.16 & --05:34:52.8 & 5.710 & 99.0 & 99.0 & 29.8 & 99.0 & 27.9 & 25.3 & NB816 &  & Ji18, Ha19, Ni20 \\
HSC J021736--052701 & 02:17:36.38 & --05:27:01.7 & 5.674 & 99.0 & 29.4 & 26.8 & 26.4 & 26.5 & 25.3 & NB816 &  & Ji17, Ji18, Ha19, Ni20 \\
HSC J021736--053027 & 02:17:36.68 & --05:30:27.6 & 5.686 & 99.0 & 99.0 & 27.9 & 27.2 & 99.0 & 25.7 & NB816 &  & Ha19 \\
HSC J021737--043943 & 02:17:37.97 & --04:39:43.1 & 5.755 & 29.2 & 99.0 & 27.0 & 26.5 & 26.1 & 24.7 & NB816 &  & On21 \\
HSC J021738--053048 & 02:17:38.29 & --05:30:48.8 & 5.687 & 99.0 & 29.0 & 28.8 & 27.9 & 27.8 & 25.9 & NB816 &  & Ha19 \\
HSC J021739--043837 & 02:17:39.26 & --04:38:37.4 & 5.722 & 99.0 & 99.0 & 27.9 & 27.6 & 99.0 & 24.8 & NB816 &  & Ni20, On21 \\
HSC J021740--044544 & 02:17:40.60 & --04:45:44.2 & 5.704 & 99.0 & 99.0 & 99.0 & 99.0 & 99.0 & 26.0 & NB816 &  & This Study \\
HSC J021741--044426 & 02:17:41.29 & --04:44:26.7 & 5.704 & 99.0 & 99.0 & 27.9 & 27.5 & 28.2 & 25.8 & NB816 &  & Ni20, This Study \\
HSC J021742--052810 & 02:17:42.18 & --05:28:10.6 & 5.679 & 28.8 & 28.5 & 27.0 & 26.3 & 26.0 & 25.5 & NB816 &  & Ha19, Ni20 \\
HSC J021745--044129 & 02:17:45.74 & --04:41:29.3 & 5.676 & 29.3 & 99.0 & 26.8 & 27.5 & 26.6 & 25.1 & NB816 &  & Ni20, On21 \\
HSC J021745--052936 & 02:17:45.26 & --05:29:36.2 & 5.688 & 29.4 & 99.0 & 26.7 & 26.2 & 26.6 & 24.6 & NB816 &  & Ou08, Ji17, Ji18, Ni20 \\
HSC J021747--052252 & 02:17:47.42 & --05:22:52.0 & 5.722 & 99.0 & 99.0 & 28.8 & 28.1 & 99.0 & 25.7 & NB816 &  & This Study \\
HSC J021748--043948 & 02:17:48.67 & --04:39:48.9 & 5.668 & 99.0 & 99.0 & 27.8 & 27.6 & 27.5 & 26.0 & NB816 &  & Ni20, This Study \\
HSC J021748--051550 & 02:17:48.77 & --05:15:50.0 & 5.681 & 99.0 & 99.0 & 27.2 & 26.8 & 99.0 & 25.9 & NB816 &  & Ji18, Ni20 \\
HSC J021748--053127 & 02:17:48.47 & --05:31:27.1 & 5.690 & 99.0 & 29.9 & 26.9 & 26.3 & 26.7 & 24.6 & NB816 &  & Ou08, Ji17, Ji18, Ni20 \\
HSC J021750--052708 & 02:17:50.00 & --05:27:08.2 & 5.693 & 99.0 & 99.0 & 27.3 & 26.8 & 26.2 & 25.0 & NB816 &  & Ou08, Ji18, Ni20 \\
HSC J021751--053003 & 02:17:51.15 & --05:30:03.8 & 5.712 & 99.0 & 28.6 & 27.2 & 26.3 & 26.2 & 25.3 & NB816 &  & Ou08, Ji18, Ni20 \\
HSC J021755--043251 & 02:17:55.40 & --04:32:51.6 & 5.692 & 28.9 & 28.6 & 27.3 & 26.7 & 27.1 & 24.9 & NB816 &  & Ni20, On21 \\
HSC J021755--053027 & 02:17:55.84 & --05:30:27.1 & 5.694 & 99.0 & 99.0 & 27.9 & 27.1 & 99.0 & 25.7 & NB816 &  & Ha19 \\
HSC J021757--053309 & 02:17:57.67 & --05:33:09.5 & 5.749 & 29.8 & 99.0 & 27.0 & 26.5 & 26.9 & 24.9 & NB816 &  & Ji18, Ha19, Ni20 \\
HSC J021758--043030 & 02:17:58.92 & --04:30:30.5 & 5.690 & 29.7 & 99.0 & 26.6 & 25.9 & 26.0 & 24.5 & NB816 &  & Ji17, Ni20 \\
HSC J021759--042758 & 02:17:59.91 & --04:27:58.0 & 5.661 & 99.0 & 99.0 & 27.4 & 26.9 & 27.1 & 26.1 & NB816 &  & Ni20, This Study \\
HSC J021759--045346 & 02:17:59.00 & --04:53:46.0 & 5.682 & 28.9 & 28.1 & 27.3 & 27.5 & 26.7 & 25.6 & NB816 &  & Ni20 \\
HSC J021759--052015 & 02:17:59.76 & --05:20:15.1 & 5.700 & 99.0 & 99.0 & 28.0 & 27.9 & 28.9 & 25.4 & NB816 &  & Ni20 \\
HSC J021800--053519 & 02:18:00.73 & --05:35:19.2 & 5.673 & 99.0 & 99.0 & 26.6 & 25.3 & 25.3 & 25.0 & NB816 &  & Ha19, Ni20 \\
HSC J021802--050038 & 02:18:02.30 & --05:00:38.8 & 5.733 & 99.0 & 99.0 & 28.3 & 29.4 & 99.0 & 25.4 & NB816 &  & Ni20 \\
HSC J021802--052011 & 02:18:02.19 & --05:20:11.6 & 5.715 & 99.0 & 99.0 & 28.1 & 27.6 & 28.2 & 25.4 & NB816 &  & Ji18, Ha19, Ni20 \\
HSC J021803--044653 & 02:18:03.59 & --04:46:53.2 & 5.670 & 99.0 & 28.4 & 27.8 & 27.3 & 27.7 & 25.7 & NB816 &  & Ni20, This Study \\
HSC J021803--051356 & 02:18:03.81 & --05:13:56.6 & 5.697 & 99.0 & 99.0 & 27.8 & 27.5 & 27.3 & 25.2 & NB816 &  & Ni20 \\
HSC J021803--052643 & 02:18:03.89 & --05:26:43.5 & 5.747 & 99.0 & 29.2 & 27.2 & 26.9 & 27.0 & 24.8 & NB816 &  & Ji18, Ha19, Ni20 \\
HSC J021804--052147 & 02:18:04.18 & --05:21:47.3 & 5.734 & 99.0 & 29.1 & 26.7 & 25.8 & 25.8 & 24.7 & NB816 &  & Ji18, Ha19, Ni20 \\
HSC J021805--052027 & 02:18:05.29 & --05:20:27.1 & 5.742 & 99.0 & 99.0 & 27.9 & 27.0 & 27.4 & 25.0 & NB816 &  & Ji18, Ha19, Ni20 \\
HSC J021805--052704 & 02:18:05.19 & --05:27:04.2 & 5.748 & 99.0 & 99.0 & 27.4 & 27.4 & 26.6 & 24.9 & NB816 &  & Ji18, Ha19, Ni20 \\
HSC J021806--042847 & 02:18:06.16 & --04:28:47.7 & 5.727 & 28.9 & 29.4 & 27.3 & 27.0 & 27.0 & 25.4 & NB816 &  & Ni20, On21 \\
HSC J021807--043058 & 02:18:07.06 & --04:30:58.4 & 5.725 & 28.9 & 28.3 & 27.1 & 27.1 & 28.0 & 25.5 & NB816 &  & Ni20 \\
HSC J021807--051526 & 02:18:07.70 & --05:15:26.8 & 5.750 & 99.0 & 99.0 & 26.8 & 26.5 & 26.8 & 25.0 & NB816 &  & Ji18, Ni20 \\
HSC J021810--053708 & 02:18:10.68 & --05:37:08.0 & 5.749 & 99.0 & 28.9 & 26.9 & 26.5 & 26.2 & 24.4 & NB816 &  & Ji18, Ni20 \\
HSC J021814--053249 & 02:18:14.40 & --05:32:49.2 & 5.673 & 99.0 & 29.1 & 26.6 & 26.0 & 25.5 & 25.2 & NB816 &  & Ou08, Ji18, Ni20 \\
HSC J021819--050355 & 02:18:19.58 & --05:03:55.5 & 5.743 & 99.0 & 29.8 & 26.9 & 26.2 & 26.1 & 25.0 & NB816 &  & Ni20, On21 \\
HSC J021821--044900 & 02:18:21.28 & --04:49:00.3 & 5.705 & 29.0 & 29.4 & 27.5 & 26.9 & 26.6 & 25.3 & NB816 &  & Ni20 \\
HSC J021822--042926 & 02:18:22.92 & --04:29:26.0 & 5.697 & 99.0 & 99.0 & 28.6 & 99.0 & 99.0 & 25.3 & NB816 &  & On21 \\
HSC J021823--044335 & 02:18:23.29 & --04:43:35.1 & 5.670 & 99.0 & 29.4 & 26.1 & 25.0 & 25.1 & 24.8 & NB816 &  & Ji17, Ni20 \\
HSC J021823--053205 & 02:18:23.82 & --05:32:05.4 & 5.679 & 29.2 & 29.8 & 27.4 & 27.5 & 27.6 & 25.8 & NB816 &  & Ou08, Ji18, Ni20 \\
HSC J021825--050700 & 02:18:25.43 & --05:07:00.6 & 5.683 & 29.3 & 99.0 & 29.3 & 27.5 & 99.0 & 25.6 & NB816 &  & On21 \\
HSC J021827--044737 & 02:18:27.45 & --04:47:37.1 & 5.703 & 99.0 & 99.0 & 26.4 & 26.4 & 26.5 & 23.9 & NB816 &  & Ji17, Ni20 \\
HSC J021828--051423 & 02:18:28.89 & --05:14:23.1 & 5.737 & 99.0 & 99.0 & 26.7 & 26.6 & 26.2 & 24.2 & NB816 &  & Ha19 \\
HSC J021830--051457 & 02:18:30.54 & --05:14:57.8 & 5.688 & 99.0 & 29.7 & 26.3 & 26.6 & 26.8 & 23.9 & NB816 &  & Ha19 \\
HSC J021835--042321 & 02:18:35.95 & --04:23:21.7 & 5.755 & 99.0 & 99.0 & 25.6 & 25.3 & 25.6 & 23.4 & NB816 &  & Sh18 \\
HSC J021836--053208 & 02:18:36.15 & --05:32:08.8 & 5.726 & 99.0 & 29.9 & 28.4 & 99.0 & 27.3 & 25.1 & NB816 &  & Ni20 \\
HSC J021836--053528 & 02:18:36.38 & --05:35:28.1 & 5.698 & 29.1 & 29.2 & 26.3 & 25.7 & 25.3 & 23.9 & NB816 &  & On21 \\
HSC J021845--052915 & 02:18:45.03 & --05:29:15.9 & 5.660 & 99.0 & 29.6 & 26.5 & 26.0 & 25.8 & 25.2 & NB816 &  & On21 \\
HSC J021846--043722 & 02:18:46.66 & --04:37:22.4 & 5.735 & 99.0 & 29.5 & 27.6 & 27.3 & 26.8 & 25.0 & NB816 &  & On21 \\
HSC J021847--052725 & 02:18:47.77 & --05:27:25.0 & 5.662 & 99.0 & 99.0 & 27.7 & 27.0 & 26.8 & 25.8 & NB816 &  & This Study \\
HSC J021848--051715 & 02:18:48.23 & --05:17:15.6 & 5.741 & 29.3 & 99.0 & 26.9 & 25.9 & 25.7 & 24.7 & NB816 &  & On21 \\
HSC J021848--052235 & 02:18:48.99 & --05:22:35.4 & 5.718 & 29.7 & 28.6 & 26.6 & 26.2 & 26.4 & 24.8 & NB816 &  & On21 \\
HSC J021850--053007 & 02:18:50.99 & --05:30:07.6 & 5.705 & 99.0 & 28.3 & 27.2 & 25.9 & 25.6 & 25.0 & NB816 &  & On21 \\
HSC J021855--044022 & 02:18:55.47 & --04:40:22.8 & 5.659 & 29.0 & 29.4 & 27.4 & 27.2 & 29.1 & 26.1 & NB816 &  & This Study \\
HSC J021857--045648 & 02:18:57.33 & --04:56:48.9 & 5.681 & 29.7 & 99.0 & 28.2 & 27.1 & 99.0 & 25.4 & NB816 &  & Ni20, On21 \\
HSC J021859--052916 & 02:18:59.93 & --05:29:16.8 & 5.674 & 29.1 & 27.8 & 25.6 & 24.8 & 24.7 & 24.3 & NB816 &  & On21 \\
HSC J021911--045707 & 02:19:11.04 & --04:57:07.6 & 5.704 & 99.0 & 99.0 & 27.7 & 28.7 & 28.6 & 25.2 & NB816 &  & Ni20, On21 \\
HSC J021916--051402 & 02:19:16.44 & --05:14:02.9 & 5.710 & 99.0 & 29.2 & 26.7 & 26.3 & 26.2 & 24.8 & NB816 &  & Ni20 \\
HSC J021943--044914 & 02:19:43.91 & --04:49:14.4 & 5.684 & 99.0 & 99.0 & 27.9 & 27.2 & 26.8 & 25.4 & NB816 &  & Ni20, On21 \\
HSC J021945--044659 & 02:19:45.86 & --04:46:59.5 & 5.691 & 29.6 & 99.0 & 27.7 & 26.9 & 26.8 & 25.6 & NB816 &  & Ni20 \\
HSC J021951--051038 & 02:19:51.16 & --05:10:38.4 & 5.667 & 99.0 & 29.2 & 26.9 & 27.1 & 26.5 & 25.6 & NB816 &  & Ni20 \\
HSC J022001--045839 & 02:20:01.06 & --04:58:39.7 & 5.682 & 99.0 & 99.0 & 27.1 & 27.2 & 27.1 & 25.5 & NB816 &  & Ni20 \\
HSC J022001--045839 & 02:20:01.06 & --04:58:39.7 & 5.682 & 99.0 & 99.0 & 27.1 & 27.2 & 27.1 & 25.5 & NB816 &  & On21 \\
HSC J022001--051637 & 02:20:01.11 & --05:16:37.4 & 5.711 & 29.7 & 29.2 & 26.1 & 25.7 & 25.8 & 23.6 & NB816 &  & Sh18 \\
HSC J022003--045416 & 02:20:03.20 & --04:54:16.6 & 5.697 & 99.0 & 29.5 & 26.3 & 25.3 & 25.2 & 24.6 & NB816 &  & On21 \\
HSC J022012--044950 & 02:20:12.14 & --04:49:50.9 & 5.681 & 99.0 & 99.0 & 27.0 & 26.7 & 28.6 & 25.3 & NB816 &  & Ou08, Ni20 \\
HSC J022017--045046 & 02:20:17.68 & --04:50:46.7 & 5.675 & 99.0 & 99.0 & 27.9 & 27.6 & 99.0 & 26.1 & NB816 &  & This Study \\
HSC J022021--045315 & 02:20:21.50 & --04:53:15.2 & 5.671 & 99.0 & 28.4 & 27.2 & 27.1 & 28.3 & 25.9 & NB816 &  & Ou08, Ni20 \\
HSC J022026--045217 & 02:20:26.89 & --04:52:17.8 & 5.720 & 28.9 & 27.7 & 26.3 & 25.8 & 25.6 & 25.1 & NB816 &  & Ni20, On21 \\
HSC J022036--045819 & 02:20:36.28 & --04:58:19.5 & 5.699 & 99.0 & 99.0 & 27.7 & 27.6 & 99.0 & 25.4 & NB816 &  & Ni20 \\
HSC J095835+014212 & 09:58:35.63 & +01:42:12.8 & 5.755 & 99.0 & 99.0 & 26.8 & 26.6 & 26.5 & 24.6 & NB816 &  & Ni20 \\
HSC J095841+015027 & 09:58:41.83 & +01:50:27.7 & 5.705 & 99.0 & 99.0 & 27.6 & 27.4 & 28.4 & 25.6 & NB816 &  & Ni20 \\
HSC J095850+014148 & 09:58:50.72 & +01:41:48.9 & 5.700 & 99.0 & 99.0 & 27.4 & 26.5 & 27.4 & 24.9 & NB816 &  & Ni20 \\
HSC J095905+014747 & 09:59:05.40 & +01:47:47.8 & 5.685 & 99.0 & 99.0 & 27.5 & 27.2 & 27.7 & 25.1 & NB816 &  & Ni20 \\
HSC J095908+020847 & 09:59:08.42 & +02:08:47.1 & 5.689 & 99.0 & 29.9 & 27.9 & 26.9 & 26.9 & 25.9 & NB816 &  & Ni20 \\
HSC J095911+015237 & 09:59:11.28 & +01:52:37.4 & 5.678 & 99.0 & 99.0 & 27.2 & 26.3 & 26.2 & 25.7 & NB816 &  & Ni20 \\
HSC J095912+014436 & 09:59:12.14 & +01:44:36.1 & 5.686 & 99.0 & 99.0 & 28.3 & 28.0 & 99.0 & 26.1 & NB816 &  & Ni20 \\
HSC J095918+020912 & 09:59:18.94 & +02:09:12.1 & 5.729 & 99.0 & 99.0 & 27.9 & 27.8 & 27.4 & 25.3 & NB816 &  & Ni20 \\
HSC J095919+020321 & 09:59:19.74 & +02:03:21.9 & 5.704 & 99.0 & 99.0 & 28.4 & 28.7 & 28.0 & 24.9 & NB816 &  & Ma12, Has18 \\
HSC J095919+020321 & 09:59:19.74 & +02:03:21.9 & 5.710 & 99.0 & 99.0 & 28.4 & 28.7 & 28.0 & 24.9 & NB816 &  & Ni20 \\
HSC J095921+014141 & 09:59:21.11 & +01:41:41.0 & 5.716 & 99.0 & 99.0 & 27.5 & 28.0 & 28.1 & 24.8 & NB816 &  & Ni20 \\
HSC J095922+021029 & 09:59:22.27 & +02:10:29.2 & 5.716 & 28.8 & 28.8 & 27.0 & 28.3 & 27.2 & 25.2 & NB816 &  & On21 \\
HSC J095923+020214 & 09:59:23.66 & +02:02:14.4 & 5.724 & 99.0 & 99.0 & 28.2 & 27.3 & 27.1 & 25.5 & NB816 &  & Ni20 \\
HSC J095930+023618 & 09:59:30.92 & +02:36:18.2 & 5.700 & 99.0 & 99.0 & 27.3 & 26.9 & 26.8 & 25.1 & NB816 &  & Ni20 \\
HSC J095932+014645 & 09:59:32.85 & +01:46:45.1 & 5.735 & 99.0 & 29.2 & 27.6 & 28.4 & 28.5 & 25.3 & NB816 &  & Ni20 \\
HSC J095940+020520 & 09:59:40.47 & +02:05:20.6 & 5.748 & 29.0 & 30.0 & 27.3 & 28.0 & 27.7 & 25.0 & NB816 &  & Ni20 \\
HSC J095944+020050 & 09:59:44.06 & +02:00:50.6 & 5.688 & 29.1 & 28.3 & 26.7 & 26.1 & 26.7 & 24.8 & NB816 &  & Ma12, Has18, Ni20 \\
HSC J095946+013208 & 09:59:46.72 & +01:32:08.3 & 5.735 & 99.0 & 99.0 & 26.5 & 25.5 & 25.7 & 24.4 & NB816 &  & Has18 \\
HSC J095946+014352 & 09:59:46.14 & +01:43:52.8 & 5.717 & 99.0 & 28.4 & 26.9 & 26.7 & 26.6 & 25.1 & NB816 &  & Ma12, Has18 \\
HSC J095947+014449 & 09:59:47.55 & +01:44:49.4 & 5.700 & 99.0 & 99.0 & 27.4 & 27.3 & 26.4 & 25.2 & NB816 &  & Ni20 \\
HSC J095950+020310 & 09:59:50.75 & +02:03:10.6 & 5.690 & 99.0 & 28.9 & 27.6 & 26.8 & 26.4 & 25.4 & NB816 &  & Ni20 \\
HSC J095950+023223 & 09:59:50.99 & +02:32:23.1 & 5.666 & 28.9 & 28.4 & 26.8 & 26.5 & 27.1 & 25.5 & NB816 &  & Ma12, Has18 \\
HSC J095952+013723 & 09:59:52.13 & +01:37:23.2 & 5.724 & 99.0 & 29.8 & 26.5 & 26.0 & 26.0 & 24.2 & NB816 &  & Ma12, Has18 \\
HSC J095952+015005 & 09:59:52.03 & +01:50:05.7 & 5.744 & 99.0 & 29.1 & 26.3 & 25.4 & 25.3 & 24.5 & NB816 &  & Ma12, Has18, Ni20 \\
HSC J095952+015700 & 09:59:52.60 & +01:57:00.7 & 5.728 & 99.0 & 29.2 & 27.0 & 27.1 & 26.5 & 25.1 & NB816 &  & Ni20 \\
HSC J095953+020705 & 09:59:53.25 & +02:07:05.3 & 5.692 & 28.8 & 28.2 & 25.7 & 24.6 & 24.5 & 24.0 & NB816 &  & Ma12, Has18 \\
HSC J095954+021039 & 09:59:54.78 & +02:10:39.2 & 5.662 & 99.0 & 29.9 & 26.8 & 26.6 & 26.4 & 25.0 & NB816 &  & Ma12, Ji17, Has18, Ni20 \\
HSC J095954+021516 & 09:59:54.52 & +02:15:16.5 & 5.688 & 99.0 & 99.0 & 27.3 & 26.8 & 27.6 & 24.8 & NB816 &  & Ma12, Has18, Ni20 \\
HSC J095955+014720 & 09:59:55.00 & +01:47:20.7 & 5.715 & 99.0 & 99.0 & 26.7 & 25.9 & 25.8 & 24.3 & NB816 &  & Ma12, Has18 \\
HSC J100001+020620 & 10:00:01.62 & +02:06:20.3 & 5.706 & 99.0 & 28.7 & 27.2 & 26.5 & 26.3 & 25.1 & NB816 &  & Ni20 \\
HSC J100005+020716 & 10:00:05.06 & +02:07:16.9 & 5.704 & 99.0 & 99.0 & 27.8 & 27.2 & 27.4 & 24.8 & NB816 &  & Ma12, Has18, Ni20 \\
HSC J100012+021930 & 10:00:12.80 & +02:19:30.9 & 5.750 & 99.0 & 99.0 & 27.6 & 26.9 & 27.3 & 25.4 & NB816 &  & Ni20 \\
HSC J100013+014026 & 10:00:13.42 & +01:40:26.3 & 5.702 & 99.0 & 28.1 & 26.8 & 25.9 & 25.7 & 25.1 & NB816 &  & Ni20 \\
HSC J100019+020103 & 10:00:19.98 & +02:01:03.2 & 5.645 & 99.0 & 99.0 & 27.0 & 26.9 & 26.5 & 25.6 & NB816 &  & Ma12, Has18, Ni20 \\
HSC J100023+015608 & 10:00:23.22 & +01:56:08.1 & 5.717 & 99.0 & 99.0 & 27.5 & 26.6 & 25.9 & 25.1 & NB816 &  & Ni20 \\
HSC J100029+015000 & 10:00:29.58 & +01:50:00.6 & 5.707 & 99.0 & 99.0 & 27.3 & 26.6 & 27.0 & 25.2 & NB816 &  & Ma12, Has18 \\
HSC J100029+024115 & 10:00:29.12 & +02:41:15.7 & 5.735 & 99.0 & 28.3 & 26.8 & 26.7 & 27.3 & 25.0 & NB816 &  & Ma12, Has18 \\
HSC J100030+021714 & 10:00:30.40 & +02:17:14.8 & 5.695 & 99.0 & 99.0 & 27.5 & 26.8 & 26.8 & 24.6 & NB816 &  & Ma12, Has18, Ni20 \\
HSC J100040+021903 & 10:00:40.21 & +02:19:03.3 & 5.719 & 29.1 & 28.5 & 26.8 & 26.5 & 26.9 & 24.9 & NB816 &  & Ma12, Ta17, Has18, Ni20 \\
HSC J100042+022135 & 10:00:42.75 & +02:21:35.7 & 5.684 & 99.0 & 99.0 & 99.0 & 99.0 & 99.0 & 26.1 & NB816 &  & Ni20 \\
HSC J100044+022719 & 10:00:44.49 & +02:27:19.1 & 5.684 & 99.0 & 99.0 & 27.3 & 27.5 & 99.0 & 25.2 & NB816 &  & Ni20 \\
HSC J100058+013642 & 10:00:58.41 & +01:36:42.8 & 5.688 & 99.0 & 28.2 & 27.1 & 27.6 & 26.3 & 25.2 & NB816 &  & Ma12, Has18 \\
HSC J100100+022926 & 10:01:00.36 & +02:29:26.3 & 5.724 & 29.5 & 29.3 & 26.8 & 26.2 & 26.4 & 24.8 & NB816 &  & Ni20 \\
HSC J100102+015144 & 10:01:02.95 & +01:51:44.7 & 5.666 & 29.7 & 28.1 & 26.6 & 26.1 & 26.1 & 25.2 & NB816 &  & Ma12, Has18, Ni20 \\
HSC J100103+021444 & 10:01:03.24 & +02:14:44.9 & 5.685 & 99.0 & 99.0 & 29.0 & 27.7 & 28.1 & 26.1 & NB816 &  & Ni20 \\
HSC J100107+015222 & 10:01:07.35 & +01:52:22.7 & 5.668 & 99.0 & 99.0 & 27.4 & 27.0 & 28.5 & 25.6 & NB816 &  & Ma12, Has18, Ni20 \\
HSC J100110+021357 & 10:01:10.16 & +02:13:57.9 & 5.682 & 99.0 & 99.0 & 99.0 & 29.2 & 99.0 & 25.5 & NB816 &  & Ni20 \\
HSC J100110+022829 & 10:01:10.06 & +02:28:29.1 & 5.681 & 29.6 & 27.8 & 26.1 & 25.0 & 25.0 & 24.5 & NB816 &  & Ma12, Has18, Ni20 \\
HSC J100123+015600 & 10:01:23.84 & +01:56:00.3 & 5.726 & 99.0 & 99.0 & 26.5 & 25.9 & 26.0 & 24.0 & NB816 &  & Ma12, Has18 \\
HSC J100126+014430 & 10:01:26.89 & +01:44:30.1 & 5.686 & 99.0 & 99.0 & 26.7 & 26.2 & 26.0 & 24.6 & NB816 &  & Ma12, Has18 \\
HSC J100127+023005 & 10:01:27.76 & +02:30:05.9 & 5.696 & 29.5 & 29.1 & 26.6 & 25.9 & 26.1 & 24.6 & NB816 &  & Ma12, Has18 \\
HSC J100129+014929 & 10:01:29.08 & +01:49:29.8 & 5.707 & 99.0 & 99.0 & 26.1 & 25.6 & 25.5 & 23.8 & NB816 &  & Ma12, Has18, Ni20 \\
HSC J100131+014320 & 10:01:31.12 & +01:43:20.3 & 5.728 & 28.9 & 28.1 & 26.7 & 26.4 & 26.4 & 24.8 & NB816 &  & Ma12, Has18 \\
HSC J100131+023105 & 10:01:31.07 & +02:31:05.9 & 5.690 & 29.5 & 99.0 & 26.9 & 26.8 & 26.8 & 24.7 & NB816 &  & Ma12, Has18 \\
HSC J100146+014827 & 10:01:46.64 & +01:48:27.1 & 5.704 & 99.0 & 28.3 & 26.9 & 26.9 & 27.3 & 24.7 & NB816 &  & On21 \\
HSC J100203+013623 & 10:02:03.15 & +01:36:23.8 & 5.742 & 29.2 & 28.1 & 26.3 & 25.9 & 26.1 & 24.2 & NB816 &  & Has18 \\
HSC J100214+021243 & 10:02:14.21 & +02:12:43.0 & 5.731 & 99.0 & 99.0 & 27.7 & 26.8 & 27.2 & 24.8 & NB816 &  & On21 \\
HSC J100301+020236 & 10:03:01.15 & +02:02:36.0 & 5.682 & 99.0 & 28.7 & 26.1 & 25.0 & 25.0 & 24.7 & NB816 &  & Ma12, Has18 \\
HSC J100305+015141 & 10:03:05.33 & +01:51:41.0 & 5.694 & 99.0 & 28.4 & 27.0 & 26.3 & 26.0 & 25.0 & NB816 &  & Ma12, Has18 \\
HSC J100306+014742 & 10:03:06.13 & +01:47:42.7 & 5.680 & 99.0 & 99.0 & 26.9 & 26.9 & 27.4 & 24.7 & NB816 &  & Ma12, Has18 \\
HSC J161403+535701 & 16:14:03.82 & +53:57:01.1 & 5.733 & 99.0 & 99.0 & 26.1 & 24.9 & 25.2 & 23.9 & NB816 & LAB & Zh20 \\
HSC J161927+551144 & 16:19:27.72 & +55:11:44.8 & 5.709 & 28.9 & 27.7 & 25.6 & 25.3 & 25.4 & 23.3 & NB816 & LAB & Zh20 \\
HSC J232558+002557 & 23:25:58.44 & +00:25:57.6 & 5.703 & 29.9 & 28.2 & 25.7 & 25.2 & 25.3 & 23.6 & NB816 &  & Sh18 \\
HSC J233408+004403 & 23:34:08.80 & +00:44:03.7 & 5.707 & 28.5 & 99.0 & 25.5 & 26.0 & 26.6 & 22.8 & NB816 &  & Sh18 \\
HSC J021650--051016 & 02:16:50.88 & --05:10:16.5 & 6.535 & 99.0 & 99.0 & 99.0 & 27.6 & 27.8 & 25.9 & NB921 &  & Ou10 \\
HSC J021653--050601 & 02:16:53.91 & --05:06:01.4 & 6.590 & 99.0 & 99.0 & 99.0 & 27.7 & 99.0 & 26.1 & NB921 &  & Ou10 \\
HSC J021654--045557 & 02:16:54.54 & --04:55:57.1 & 6.617 & 99.0 & 99.0 & 99.0 & 27.2 & 26.3 & 25.1 & NB921 &  & Ou10, Ni22 \\
HSC J021654--050004 & 02:16:54.38 & --05:00:04.3 & 6.512 & 99.0 & 99.0 & 99.0 & 26.9 & 99.0 & 25.4 & NB921 &  & Ou10, Ni22 \\
HSC J021658--045556 & 02:16:58.26 & --04:55:56.9 & 6.573 & 99.0 & 99.0 & 99.0 & 27.2 & 99.0 & 25.5 & NB921 &  & Ou10 \\
HSC J021702--050604 & 02:17:02.56 & --05:06:04.7 & 6.545 & 29.0 & 99.0 & 29.5 & 26.5 & 27.3 & 24.9 & NB921 &  & Ou10 \\
HSC J021703--045619 & 02:17:03.47 & --04:56:19.1 & 6.589 & 99.0 & 99.0 & 99.0 & 26.6 & 25.8 & 24.7 & NB921 &  & Ou10 \\
HSC J021746--042539 & 02:17:46.31 & --04:25:39.4 & 6.577 & 29.3 & 99.0 & 99.0 & 27.0 & 27.8 & 25.5 & NB921 &  & This Study \\
HSC J021751--042450 & 02:17:51.42 & --04:24:50.0 & 6.559 & 99.0 & 29.0 & 28.6 & 27.1 & 27.6 & 26.1 & NB921 &  & Ni22 \\
HSC J021757--050414 & 02:17:57.00 & --05:04:14.3 & 6.570 & 99.0 & 99.0 & 99.0 & 27.0 & 26.7 & 25.9 & NB921 &  & Ha19 \\
HSC J021757--050844 & 02:17:57.58 & --05:08:44.7 & 6.595 & 99.0 & 99.0 & 99.0 & 25.7 & 25.3 & 23.8 & NB921 & Himiko & Ou10, Ji17, Ha19, Ni22 \\
HSC J021757--051556 & 02:17:57.32 & --05:15:56.5 & 6.564 & 29.0 & 99.0 & 99.0 & 99.0 & 26.2 & 26.0 & NB921 &  & Ha19 \\
HSC J021800--050330 & 02:18:00.80 & --05:03:30.4 & 6.613 & 29.8 & 99.0 & 99.0 & 27.3 & 28.6 & 25.5 & NB921 &  & Ha19 \\
HSC J021800--050346 & 02:18:00.23 & --05:03:46.9 & 6.601 & 99.0 & 99.0 & 99.0 & 27.6 & 26.7 & 25.5 & NB921 &  & Ha19 \\
HSC J021806--042657 & 02:18:06.28 & --04:26:57.3 & 6.570 & 29.8 & 29.6 & 99.0 & 29.2 & 28.2 & 25.9 & NB921 &  & This Study \\
HSC J021819--050900 & 02:18:19.40 & --05:09:00.8 & 6.563 & 99.0 & 99.0 & 99.0 & 26.9 & 26.3 & 25.2 & NB921 &  & Ou10, Ha19 \\
HSC J021820--051109 & 02:18:20.70 & --05:11:09.9 & 6.575 & 99.0 & 99.0 & 99.0 & 27.5 & 27.6 & 25.0 & NB921 &  & Ou10, Ha19 \\
HSC J021827--050629 & 02:18:27.95 & --05:06:29.9 & 6.597 & 99.0 & 99.0 & 99.0 & 27.4 & 99.0 & 25.4 & NB921 &  & Ha19 \\
HSC J021827--050727 & 02:18:27.01 & --05:07:27.0 & 6.553 & 99.0 & 99.0 & 28.8 & 27.1 & 99.0 & 25.1 & NB921 &  & Ou10, Ha19, Ni22 \\
HSC J021834--053024 & 02:18:34.53 & --05:30:24.3 & 6.580 & 99.0 & 99.0 & 99.0 & 28.4 & 25.6 & 25.6 & NB921 &  & This Study \\
HSC J021839--042639 & 02:18:39.74 & --04:26:39.7 & 6.553 & 99.0 & 29.1 & 28.2 & 27.9 & 26.6 & 25.6 & NB921 &  & This Study \\
HSC J021842--043011 & 02:18:42.60 & --04:30:11.4 & 6.548 & 99.0 & 99.0 & 99.0 & 26.7 & 26.6 & 24.9 & NB921 &  & On21 \\
HSC J021843--050915 & 02:18:43.64 & --05:09:15.7 & 6.512 & 99.0 & 28.8 & 28.1 & 26.0 & 25.5 & 24.7 & NB921 &  & Sh18, Ha18, Ni22 \\
HSC J021844--043636 & 02:18:44.66 & --04:36:36.3 & 6.621 & 99.0 & 99.0 & 99.0 & 26.5 & 26.5 & 24.7 & NB921 &  & Ou10, Ni22 \\
HSC J021901--045859 & 02:19:01.43 & --04:58:59.0 & 6.556 & 99.0 & 28.5 & 29.4 & 26.5 & 26.3 & 24.8 & NB921 &  & Ji17, Ni22 \\
HSC J021933--050820 & 02:19:33.13 & --05:08:20.8 & 6.590 & 99.0 & 99.0 & 99.0 & 26.9 & 26.8 & 24.9 & NB921 &  & On21, Ni22 \\
HSC J022023--050648 & 02:20:23.81 & --05:06:48.9 & 6.550 & 99.0 & 99.0 & 99.0 & 27.5 & 99.0 & 25.6 & NB921 &  & Ni22, This Study \\
HSC J022026--050542 & 02:20:26.83 & --05:05:42.5 & 6.566 & 99.0 & 99.0 & 99.0 & 27.0 & 29.2 & 25.2 & NB921 &  & On21, Ni22 \\
HSC J095935+022505 & 09:59:35.07 & +02:25:05.2 & 6.573 & 99.0 & 30.0 & 28.8 & 26.6 & 26.8 & 25.0 & NB921 &  & Ni22 \\
HSC J100018+020008 & 10:00:18.35 & +02:00:08.2 & 6.574 & 99.0 & 99.0 & 99.0 & 27.7 & 26.5 & 24.9 & NB921 &  & Ni22 \\
HSC J100058+014815 & 10:00:58.00 & +01:48:15.1 & 6.604 & 99.0 & 99.0 & 29.8 & 25.5 & 24.8 & 23.5 & NB921 & LAB (CR7) & So15, Has18, Ni22 \\
HSC J100124+023145 & 10:01:24.79 & +02:31:45.5 & 6.545 & 29.2 & 29.6 & 28.6 & 25.8 & 25.9 & 24.0 & NB921 & LAB (MASOSA) & So15, Ji17, Ni22 \\
HSC J100207+023217 & 10:02:07.81 & +02:32:17.2 & 6.616 & 99.0 & 99.0 & 99.0 & 27.1 & 26.7 & 25.0 & NB921 &  & On21 \\
HSC J160107+550720 & 16:01:07.44 & +55:07:20.7 & 6.573 & 28.5 & 99.0 & 28.6 & 26.0 & 25.2 & 24.2 & NB921 &  & Sh18 \\
HSC J160707+555347 & 16:07:07.46 & +55:53:47.9 & 6.564 & 99.0 & 99.0 & 99.0 & 26.1 & 99.0 & 24.1 & NB921 &  & Sh18 \\
HSC J233125--005216 & 23:31:25.36 & --00:52:16.5 & 6.559 & 99.0 & 99.0 & 99.0 & 25.4 & 27.5 & 23.3 & NB921 &  & Sh18 \\
HSC J100205+020646 & 10:02:05.96 & +02:06:46.2 & 6.936 & 99.0 & 99.0 & 99.0 & 99.0 & 25.5 & 24.2 & NB973 &  & Hu17 \\
\enddata
\tablecomments{
(1) Object ID. (2) Right ascension. (3) Declination. (4) Spectroscopic redshift. (5)-(10) 2{\mbox{$.\!\!\arcsec$}}0-diameter circular aperture magnitude in g, r, i, z, y, and NB. (11) Detection NB. (12) Notable features, if any. (13) Reference for spectroscopic redshifts: Ou08: \citet{Ouchi2008}; Li09: \citet{Lilly2009}; Ou10: \citet{Ouchi2010}; Co11: \citet{Coil2011a}; Ma12: \citet{Mallery2012}; Mas12: \citet{Masters2012}; Le13: \citet{LeFevre2013}; Sh14: \citet{Shibuya2014}; Kr15: \citet{Kriek2015}; Li15: \citet{Liske2015}; So15: \citet{Sobral2015}; Mo16: \citet{Momcheva2016}; Ta17: \citet{Tasca2017}; Hu17: \citet{Hu2017a}; Ji17: \citet{Jiang2017}; Ha18: \citet{Harikane2018}; Has18: \citet{Hasinger2018}; Ji18: \citet{Jiang2018}; 
Sc18: \citet{Scodeggio2018}; Sh18a: \citet{Shibuya2018laelab}; Sh18b: \citet{Shibuya2018spec}; Ha19: \citet[][see also \citealt{Higuchi2019}]{Harikane2019}; Ly20: \citet{Lyke2020}; Ni20: \citet{Ning2020a};
On21: \citet{Ono2021}; Ni22: \citet{Ning2022a}.
}
\end{deluxetable*}
\end{longrotatetable}

\begin{longrotatetable}
\begin{deluxetable*}{lcccccccccccc}
\tablecaption{List of Lower-Redshift Objects in Our LAE Catalogs\label{tab:lowzspec}}
\tablehead{
\colhead{ID} & \colhead{R.A.} & \colhead{Decl.} & \colhead{$z_\mathrm{spec}$} & \colhead{$\mathrm{g_{ap}}$} & \colhead{$\mathrm{r_{ap}}$} & \colhead{$\mathrm{i_{ap}}$} & \colhead{$\mathrm{z_{ap}}$} & \colhead{$\mathrm{y_{ap}}$} & \colhead{$\mathrm{NB_{ap}}$} & \colhead{Sample} & \colhead{Feature} & \colhead{Reference}  \\ 
\colhead{(1)} & \colhead{(2)} & \colhead{(3)} & \colhead{(4)} & 
\colhead{(5)} & \colhead{(6)} & \colhead{(7)} &
\colhead{(8)} & \colhead{(9)} & \colhead{(10)} & \colhead{(11)} & \colhead{(12)} & \colhead{(13)}
} 
\startdata
HSC J022140--052310 & 02:21:40.99 & --05:23:10.1 & 0.000 & 21.3 & 21.7 & 22.2 & 22.4 & 22.5 & 20.9 & NB387 & STAR & Co11 \\
HSC J022151--035716 & 02:21:51.19 & --03:57:16.6 & 0.000 & 25.0 & 24.9 & 25.0 & 24.9 & 24.6 & 24.0 & NB387 & STAR & Co11 \\
HSC J022218--040755 & 02:22:18.77 & --04:07:55.5 & 0.000 & 20.8 & 21.1 & 21.4 & 21.6 & 21.7 & 20.5 & NB387 & STAR & Co11 \\
HSC J022930--042700 & 02:29:30.35 & --04:27:00.1 & 0.000 & 19.9 & 20.3 & 20.8 & 21.1 & 21.3 & 19.3 & NB387 & STAR & Sc18 \\
HSC J100030+023042 & 10:00:30.05 & +02:30:42.7 & 0.000 & 25.8 & 26.0 & 26.0 & 26.1 & 25.9 & 23.2 & NB387 & STAR & Mo16 \\
HSC J100113+024435 & 10:01:13.09 & +02:44:35.4 & 0.322 & 24.0 & 23.8 & 23.6 & 23.3 & 23.2 & 22.3 & NB387 &  & Has18 \\
HSC J022116--045644 & 02:21:16.39 & --04:56:44.0 & 0.047 & 23.9 & 23.8 & 23.8 & 23.7 & 23.7 & 22.8 & NB387 &  & Co11 \\
HSC J022621--043124 & 02:26:21.27 & --04:31:24.5 & 0.466 & 24.6 & 24.3 & 24.4 & 24.4 & 24.2 & 23.9 & NB387 &  & Li15 \\
HSC J022148--045039 & 02:21:48.69 & --04:50:39.3 & 0.618 & 23.8 & 23.7 & 23.5 & 23.3 & 23.2 & 21.7 & NB387 &  & Co11 \\
HSC J233328+000858\tablenotemark{a} & 23:33:28.70 & +00:08:58.7 & 0.685 & 21.7 & 21.3 & 21.5 & 21.3 & 21.3 & 20.6 & NB387 &  & Co11 \\
HSC J022203--045421 & 02:22:03.34 & --04:54:21.3 & 1.023 & 20.7 & 20.5 & 20.6 & 20.4 & 20.3 & 19.8 & NB387 & AGN & Li15, Ly20 \\
HSC J232638--000524 & 23:26:38.12 & --00:05:24.6 & 1.032 & 21.2 & 20.9 & 21.1 & 20.6 & 20.4 & 20.0 & NB387 & AGN & Ly20 \\
HSC J022240--040501 & 02:22:40.29 & --04:05:01.5 & 1.055 & 21.2 & 20.8 & 21.1 & 20.9 & 20.8 & 20.2 & NB387 & AGN & Li15, Ly20 \\
HSC J161212+540936 & 16:12:12.57 & +54:09:36.6 & 1.056 & 19.6 & 19.4 & 19.6 & 19.6 & 19.5 & 19.0 & NB387 & AGN & Ly20 \\
HSC J022120--045445 & 02:21:20.61 & --04:54:45.7 & 1.112 & 22.6 & 22.8 & 23.1 & 23.3 & 23.4 & 22.3 & NB387 &  & Co11 \\
HSC J022259--051813 & 02:22:59.67 & --05:18:13.0 & 1.115 & 22.1 & 22.4 & 22.6 & 22.6 & 22.5 & 21.7 & NB387 &  & Co11 \\
HSC J022734--042228 & 02:27:34.23 & --04:22:28.7 & 1.138 & 21.3 & 21.2 & 21.4 & 21.2 & 21.0 & 20.7 & NB387 & AGN & Ly20 \\
HSC J100059+021224 & 10:00:59.62 & +02:12:24.6 & 1.328 & 25.9 & 26.0 & 25.8 & 25.5 & 25.2 & 24.8 & NB527 &  & Has18 \\
HSC J100043+021352 & 10:00:43.20 & +02:13:52.6 & 1.400 & 23.4 & 23.6 & 23.4 & 22.8 & 23.3 & 22.2 & NB387 &  & Mo16 \\
HSC J022550--040247 & 02:25:50.98 & --04:02:47.4 & 1.448 & 21.9 & 21.7 & 21.7 & 21.6 & 21.6 & 21.1 & NB387 & AGN & Li15, Ly20 \\
HSC J233502+003727 & 23:35:02.63 & +00:37:27.2 & 1.455 & 20.3 & 19.9 & 19.9 & 20.0 & 20.0 & 19.4 & NB387 & AGN & Ly20 \\
HSC J161255+563408 & 16:12:55.05 & +56:34:08.4 & 1.465 & 21.2 & 21.1 & 21.0 & 21.0 & 21.1 & 20.3 & NB387 & AGN & Ly20 \\
HSC J233328+000858\tablenotemark{a} & 23:33:28.70 & +00:08:58.7 & 1.471 & 21.7 & 21.3 & 21.5 & 21.3 & 21.3 & 20.6 & NB387 & AGN & Ly20 \\
HSC J022134--054128 & 02:21:34.51 & --05:41:28.0 & 1.474 & 21.7 & 21.4 & 21.3 & 21.2 & 20.8 & 20.5 & NB387 & AGN & Li15, Sc18, Ly20 \\
HSC J160830+554350 & 16:08:30.17 & +55:43:50.5 & 1.474 & 19.1 & 19.0 & 19.0 & 19.2 & 19.3 & 18.5 & NB387 & AGN & Ly20 \\
HSC J160933+532239 & 16:09:33.94 & +53:22:39.8 & 1.477 & 22.8 & 22.3 & 22.2 & 22.1 & 21.9 & 21.3 & NB387 & AGN & Ly20 \\
HSC J022458--054447 & 02:24:58.25 & --05:44:47.2 & 1.479 & 21.6 & 21.3 & 21.5 & 21.3 & 21.3 & 20.3 & NB387 & AGN & Ly20 \\
HSC J161009+554149 & 16:10:09.05 & +55:41:49.6 & 1.482 & 21.2 & 21.2 & 21.0 & 21.0 & 21.0 & 20.2 & NB387 & AGN & Ly20 \\
HSC J233029+003252 & 23:30:29.82 & +00:32:52.3 & 1.485 & 22.5 & 22.1 & 21.8 & 21.8 & 21.7 & 21.2 & NB387 & AGN & Ly20 \\
HSC J022226--041313 & 02:22:26.85 & --04:13:13.5 & 1.486 & 22.5 & 22.8 & 22.6 & 22.4 & 22.4 & 21.0 & NB387 & AGN & Co11, Ly20 \\
HSC J023000--042835 & 02:30:00.95 & --04:28:35.3 & 1.486 & 19.8 & 19.5 & 19.5 & 19.5 & 19.6 & 18.8 & NB387 & AGN & Sc18, Ly20 \\
HSC J095928+021950 & 09:59:28.33 & +02:19:50.5 & 1.486 & 20.8 & 20.6 & 20.6 & 20.6 & 20.7 & 19.8 & NB387 & AGN & Co11, Has18 \\
HSC J161009+554155 & 16:10:09.56 & +55:41:55.9 & 1.486 & 21.9 & 21.7 & 21.5 & 21.5 & 21.5 & 20.8 & NB387 & AGN & Ly20 \\
HSC J022417--052413 & 02:24:17.87 & --05:24:13.4 & 1.491 & 21.2 & 21.0 & 20.9 & 20.8 & 20.9 & 20.2 & NB387 & AGN & Ly20 \\
HSC J022237--041140 & 02:22:37.89 & --04:11:40.4 & 1.496 & 21.5 & 21.4 & 21.3 & 21.2 & 21.3 & 20.5 & NB387 & AGN & Co11, Ly20 \\
HSC J022309--041732 & 02:23:09.01 & --04:17:32.3 & 1.496 & 22.8 & 22.4 & 22.2 & 22.1 & 21.7 & 21.2 & NB387 & AGN & Ly20 \\
HSC J022429--045807 & 02:24:29.12 & --04:58:07.8 & 1.496 & 20.5 & 20.2 & 20.0 & 20.1 & 20.2 & 19.4 & NB387 & AGN & Ly20 \\
HSC J232814+000315 & 23:28:14.11 & +00:03:15.0 & 1.501 & 22.1 & 21.8 & 21.6 & 21.5 & 21.4 & 21.0 & NB387 & AGN & Ly20 \\
HSC J022016--042422 & 02:20:16.34 & --04:24:22.4 & 1.503 & 23.4 & 22.8 & 22.5 & 22.4 & 22.5 & 21.8 & NB387 & AGN & Sc18, Ly20 \\
HSC J022349--035122 & 02:23:49.52 & --03:51:22.9 & 1.504 & 21.5 & 21.3 & 21.2 & 21.2 & 21.2 & 20.5 & NB387 & AGN & Ly20 \\
HSC J022528--043641 & 02:25:28.07 & --04:36:41.9 & 1.504 & 23.5 & 23.3 & 23.2 & 23.1 & 23.1 & 22.3 & NB387 & AGN & Le13, Li15 \\
HSC J022718--043134 & 02:27:18.85 & --04:31:34.0 & 1.505 & 21.4 & 21.0 & 21.0 & 20.9 & 20.9 & 20.4 & NB387 & AGN & Le13, Li15 \\
HSC J100111+023024 & 10:01:11.93 & +02:30:24.8 & 1.506 & 21.5 & 21.3 & 21.1 & 21.1 & 21.2 & 20.5 & NB387 & AGN & Li09, Co11 \\
HSC J095815+014923 & 09:58:15.50 & +01:49:23.0 & 1.508 & 20.7 & 20.3 & 20.1 & 20.1 & 20.1 & 19.5 & NB387 & AGN & Li09, Ly20 \\
HSC J022012--052005 & 02:20:12.84 & --05:20:05.6 & 1.522 & 21.6 & 21.5 & 21.2 & 21.3 & 21.2 & 20.6 & NB387 & AGN & Ly20 \\
HSC J161352+543558 & 16:13:52.07 & +54:35:58.9 & 1.525 & 20.7 & 20.9 & 20.7 & 20.9 & 21.0 & 20.1 & NB387 & AGN & Ly20 \\
HSC J161049+541840 & 16:10:49.15 & +54:18:40.5 & 1.526 & 22.0 & 21.7 & 21.5 & 21.5 & 21.6 & 20.9 & NB387 & AGN & Ly20 \\
HSC J233436--001527 & 23:34:36.27 & --00:15:27.8 & 1.528 & 23.1 & 22.9 & 22.6 & 22.8 & 22.6 & 22.1 & NB387 & AGN & Ly20 \\
HSC J022331--042557 & 02:23:31.83 & --04:25:57.9 & 1.584 & 22.5 & 22.7 & 22.3 & 22.2 & 22.1 & 21.7 & NB387 & AGN & Co11 \\
HSC J100057+023932 & 10:00:57.80 & +02:39:32.6 & 1.410 & 23.5 & 22.9 & 22.8 & 22.8 & 22.8 & 22.3 & NB527 & AGN & Li09 \\
HSC J095812+022302 & 09:58:12.40 & +02:23:02.6 & 2.322 & 25.9 & 26.0 & 25.9 & 25.6 & 25.4 & 24.7 & NB527 &  & Has18 \\
HSC J095801+014832 & 09:58:01.45 & +01:48:32.9 & 2.402 & 22.7 & 22.7 & 22.4 & 22.1 & 21.8 & 21.3 & NB527 & AGN & Li09 \\
HSC J095823+024106 & 09:58:23.56 & +02:41:06.1 & 2.402 & 23.7 & 23.8 & 23.7 & 23.5 & 23.1 & 22.8 & NB527 & AGN & Has18 \\
HSC J095914+013635 & 09:59:14.66 & +01:36:35.1 & 2.408 & 23.1 & 23.3 & 23.2 & 23.0 & 22.6 & 21.7 & NB527 & AGN & Li09 \\
\enddata
\tablecomments{
(1) Object ID. (2) Right ascension. (3) Declination. (4) Spectroscopic redshift. (5)-(10) 2{\mbox{$.\!\!\arcsec$}}0-diameter circular aperture magnitude in g, r, i, z, y, and NB. (11) Detection NB. (12) Notable features, if any. (13) Reference for spectroscopic redshifts: Li09: \citet{Lilly2009}; Co11: \citet{Coil2011a}; Le13: \citet{LeFevre2013}; Li15: \citet{Liske2015}; Mo16: \citet{Momcheva2016}; Has18: \citet{Hasinger2018}; Sc18: \citet{Scodeggio2018}; Ka19: \citet{Kashino2019fc}; Ly20: \citet{Lyke2020}.
}
\tablenotetext{a}{duplicated, but with different redshift estimates}
\end{deluxetable*}
\end{longrotatetable}


\end{document}